\documentclass[traditabstract]{aa}

\usepackage{graphicx}
\usepackage{natbib}
\usepackage{txfonts}

\def\crm{\cr\noalign{\medskip}}
\def\m@th{\mathsurround=0pt}
\def\EQM#1{\vcenter{\normalbaselines\m@th
    \ialign{${\displaystyle ##}$\hfil&&\ ${\displaystyle ##}$\hfil\crcr
    \mathstrut\crcr\noalign{\kern-\baselineskip}
    \noalign{\smallskip}
    #1\crcr\mathstrut\crcr\noalign{\kern-\baselineskip}}}}

\newcommand{\Frac}[2]{{{\displaystyle\strut#1}\over{\displaystyle\strut#2}}}
\newcommand{\php}{\phantom{+}}

\def\ImUnit{\mathbf{i}}

\def\Id{\mathrm{I}}

\def\sun{\odot}
\def\earth{\oplus}

\def\expo#1{\mathbf{e}^{#1}}

\def\deriv#1#2{\frac{\mathrm{d} #1}{\mathrm{d} #2}}

\def\dpart#1#2{\frac{\partial #1}{\partial #2}}

\def\scaled#1{\hat{#1}}
\def\tempscaled#1{\check{#1}}

\def\secul#1{\tilde{#1}}

\def\refeq#1{Eq.~(\ref{#1})}

\def\modif#1{#1}

\graphicspath{{./figures_pdf/}}

\newcommand\figa{
  \begin{figure}
    \centering
    \includegraphics[width=8cm,trim = 1cm 0cm 4cm 3cm, clip]{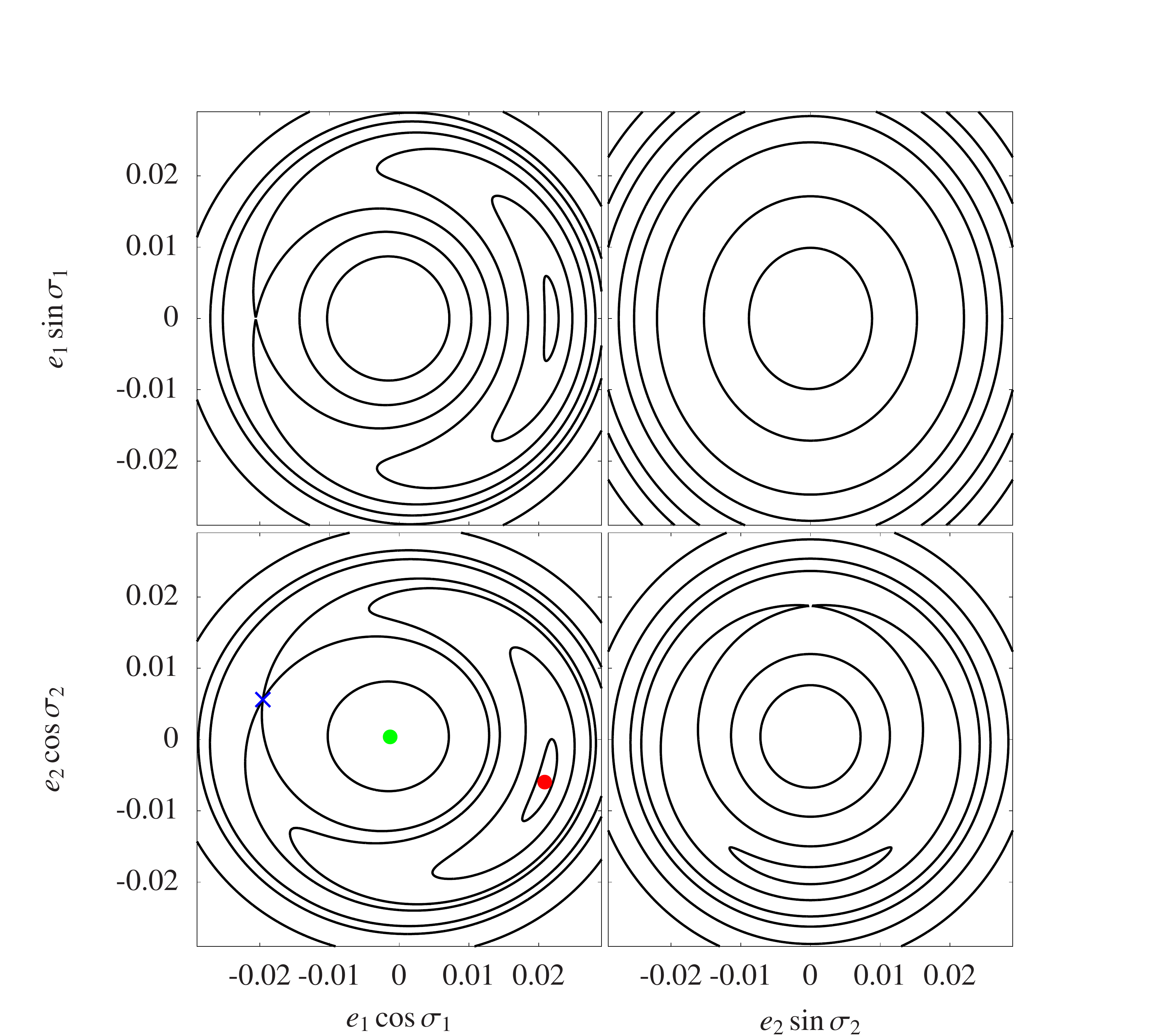}
    \caption{2:1 resonance energy levels in section planes defined by:
    $e_2=0$ (top left),
    $\cos\sigma_1=\cos\sigma_2=0$ (top right),
    $\sin\sigma_1=\sin\sigma_2=0$ (bottom left), and
    $e_1=0$ (bottom right).
    Stable ACR solutions are highlighted with colored dots (red and green).
    Unstable ones are highlighted with colored crosses (blue).
    The star mass is given by $m_0 = m_\sun$ and planets masses by $m_1=m_2=m_\earth$.
    The constant $G$ is set to $G = G_0 (1 - 10^{-4})$.
    The Hamiltonian is developed up to degree 4-1.}
    \label{Figa}
  \end{figure}
}

\newcommand\figb{
  \begin{figure*}
    \centering
    \includegraphics[width=15cm,trim = 0cm 0cm 3cm 3cm, clip]{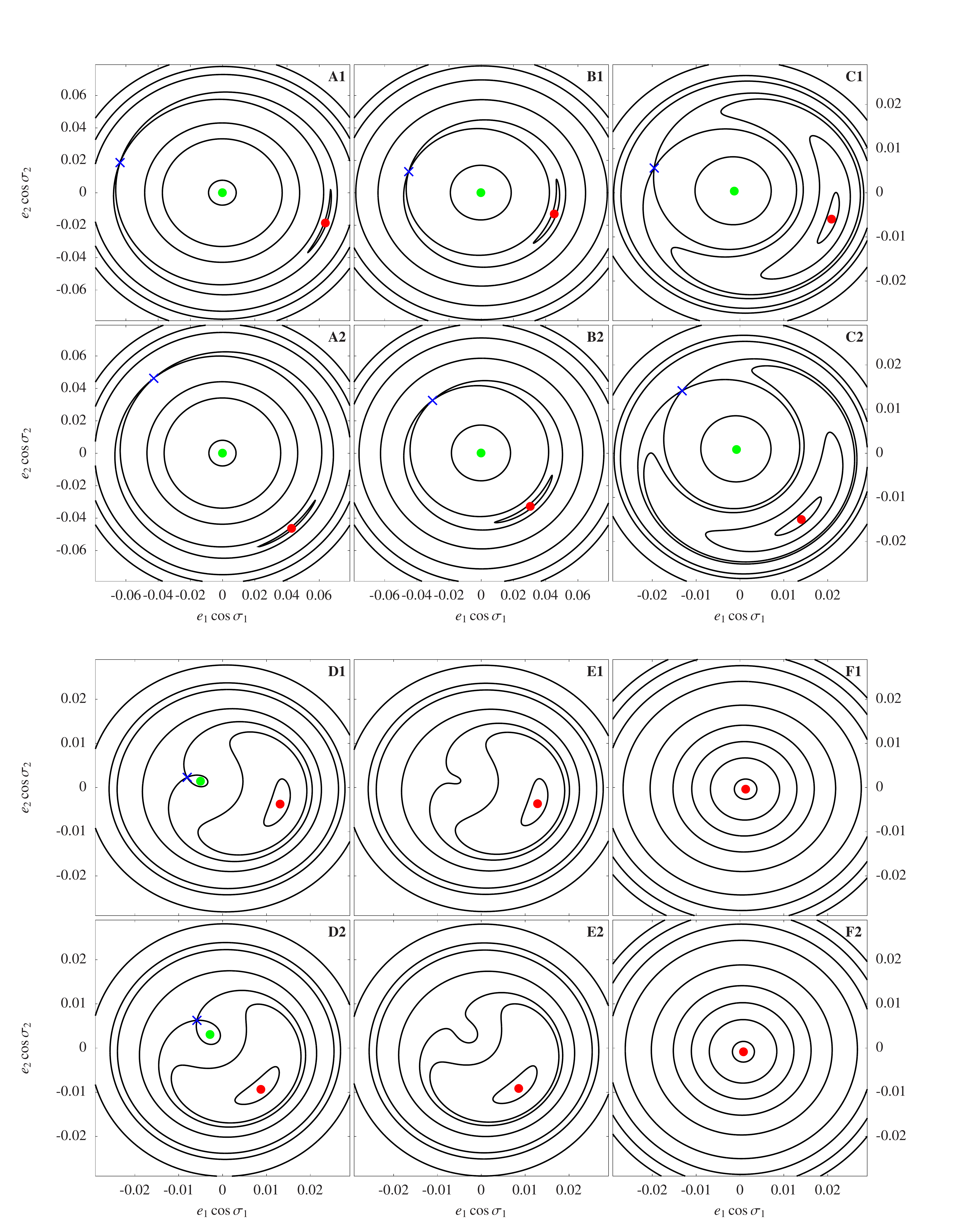}
    \caption{Energy levels sections in the plane defined by $\sin\sigma_1=\sin\sigma_2=0$
    for the 2:1 resonance (A1 to F1),
    and the 3:2 resonance (A2 to F2).
    The constant $G/G_0-1$ is set to
    $- 10^{-3}$ (A1, A2),
    $- 5\times10^{-4}$ (B1, B2),
    $- 10^{-4}$ (C1, C2),
    $- 3.2\times10^{-5}$ (D1, D2),
    $- 3\times10^{-5}$ (E1, E2), and
    $+ 10^{-3}$ (F1, F2).
    Stable ACR solutions are highlighted with colored dots (red and green).
    Unstable ones are highlighted with colored crosses (blue).
    The star mass is given by $m_0 = m_\sun$ and planets masses by $m_1=m_2=m_\earth$.
    The Hamiltonian is developed up to degree 4-1.
    Note that the scales are different for graphs A1, A2, B1, and B2 than for other graphs.}
    \label{Figb}
  \end{figure*}
}

\newcommand\figc{
  \begin{figure}
    \centering
    \includegraphics[width=8cm,trim = 0.5cm 0cm 0cm 1cm, clip]{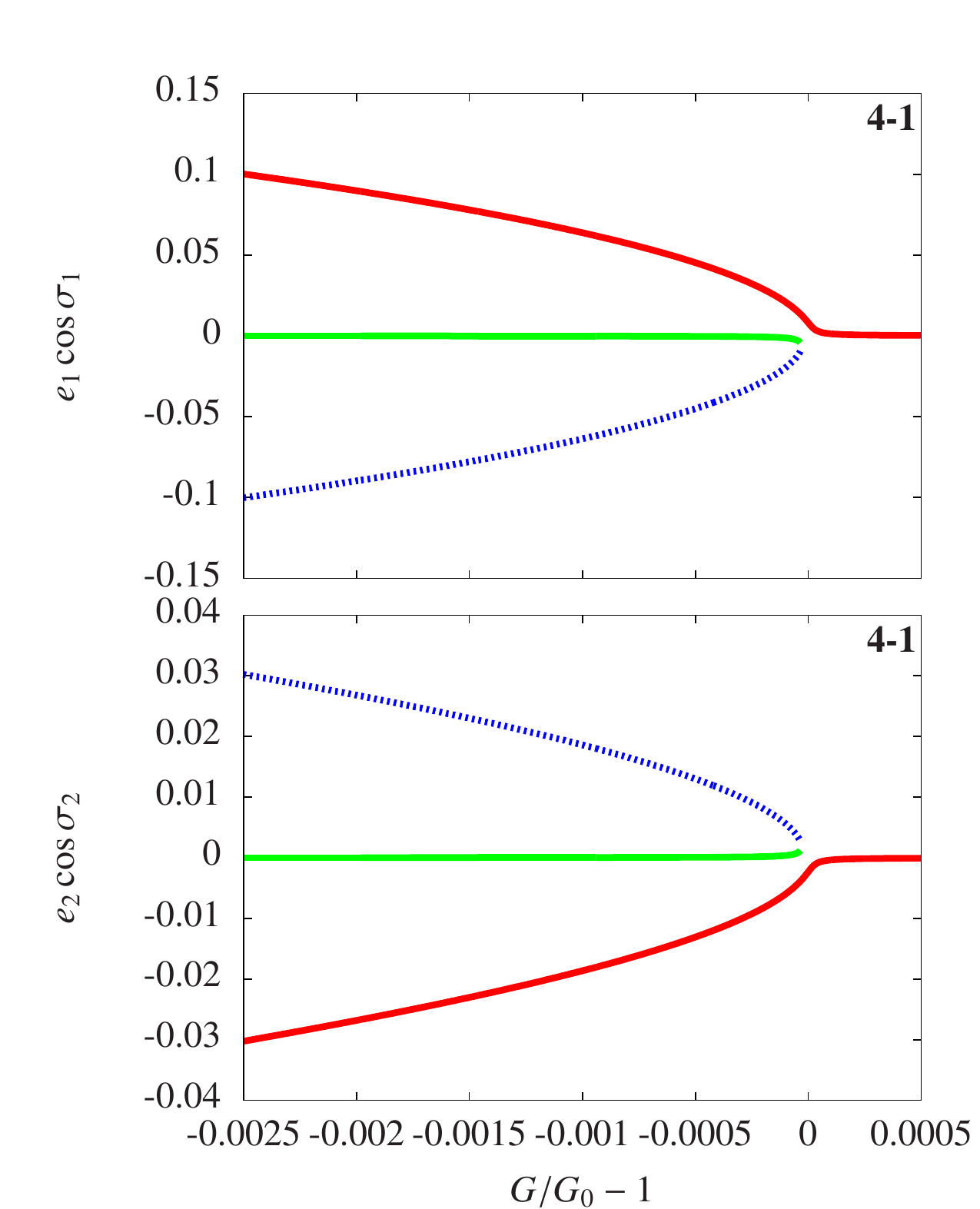}
    \caption{2:1 ACR positions as functions of $G$.
    The star mass is given by $m_0 = m_\sun$ and planets masses by $m_1=m_2=m_\earth$.
    The Hamiltonian is developed up to degree 4-1.
    Elliptic (stable) ACR are plotted using continuous lines whereas
    hyperbolic (unstable) ACR are plotted using dashed lines.
    At this degree of development, all ACR have $\sin\sigma_1=\sin\sigma_2=0$.}
    \label{Figc}
  \end{figure}
}

\newcommand\figd{
  \begin{figure}
    \centering
    \includegraphics[width=8cm,trim = 0cm 0cm 1cm 0.5cm, clip]{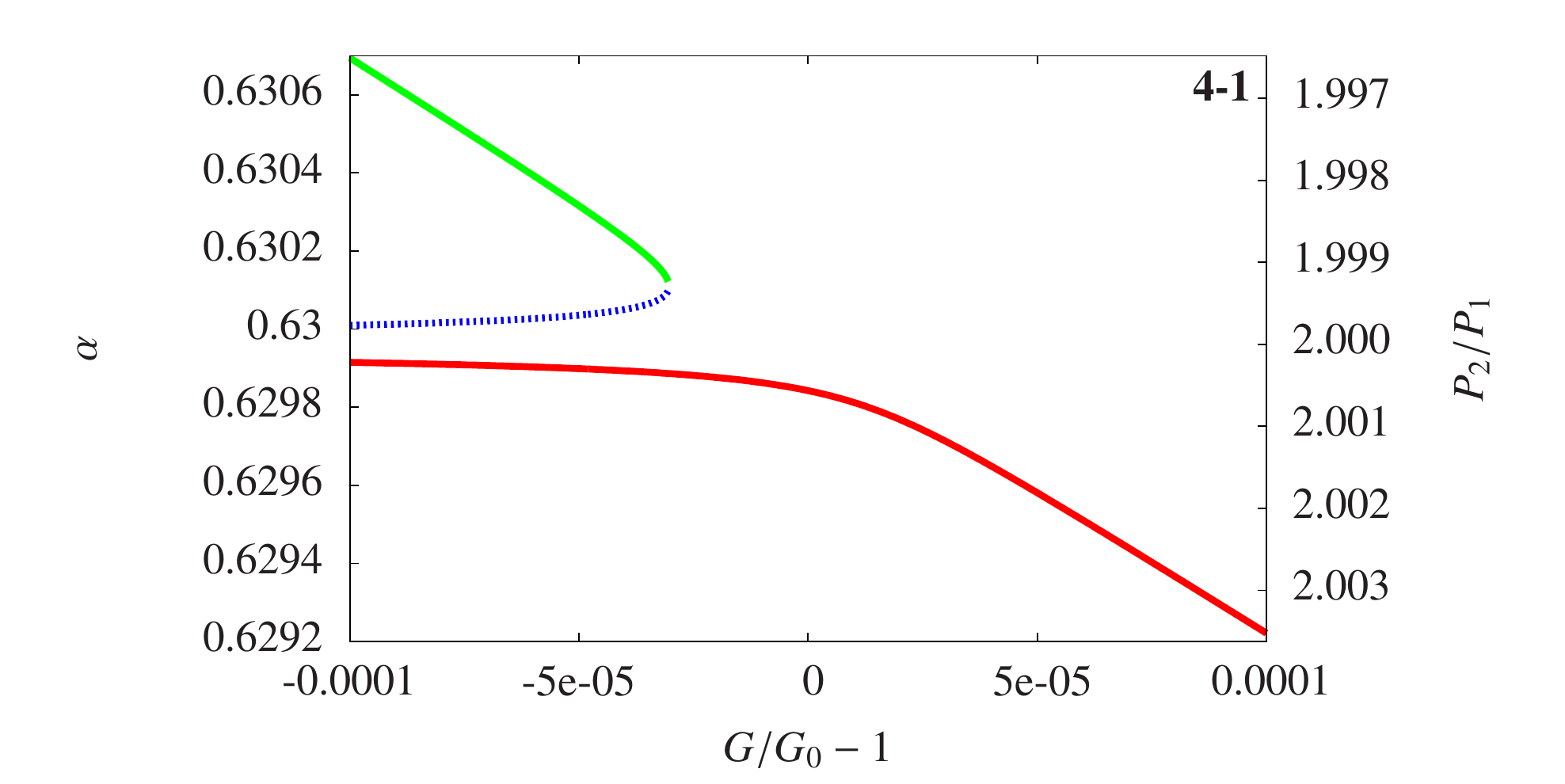}
    \caption{2:1 ACR positions in the plane $(G, \alpha)$ in the same conditions
    as for Fig.~\ref{Figc}.}
    \label{Figd}
  \end{figure}
}

\newcommand\fige{
  \begin{figure*}
    \centering
    \includegraphics[width=15cm,trim = 0.5cm 0cm 3.5cm 1.5cm, clip]{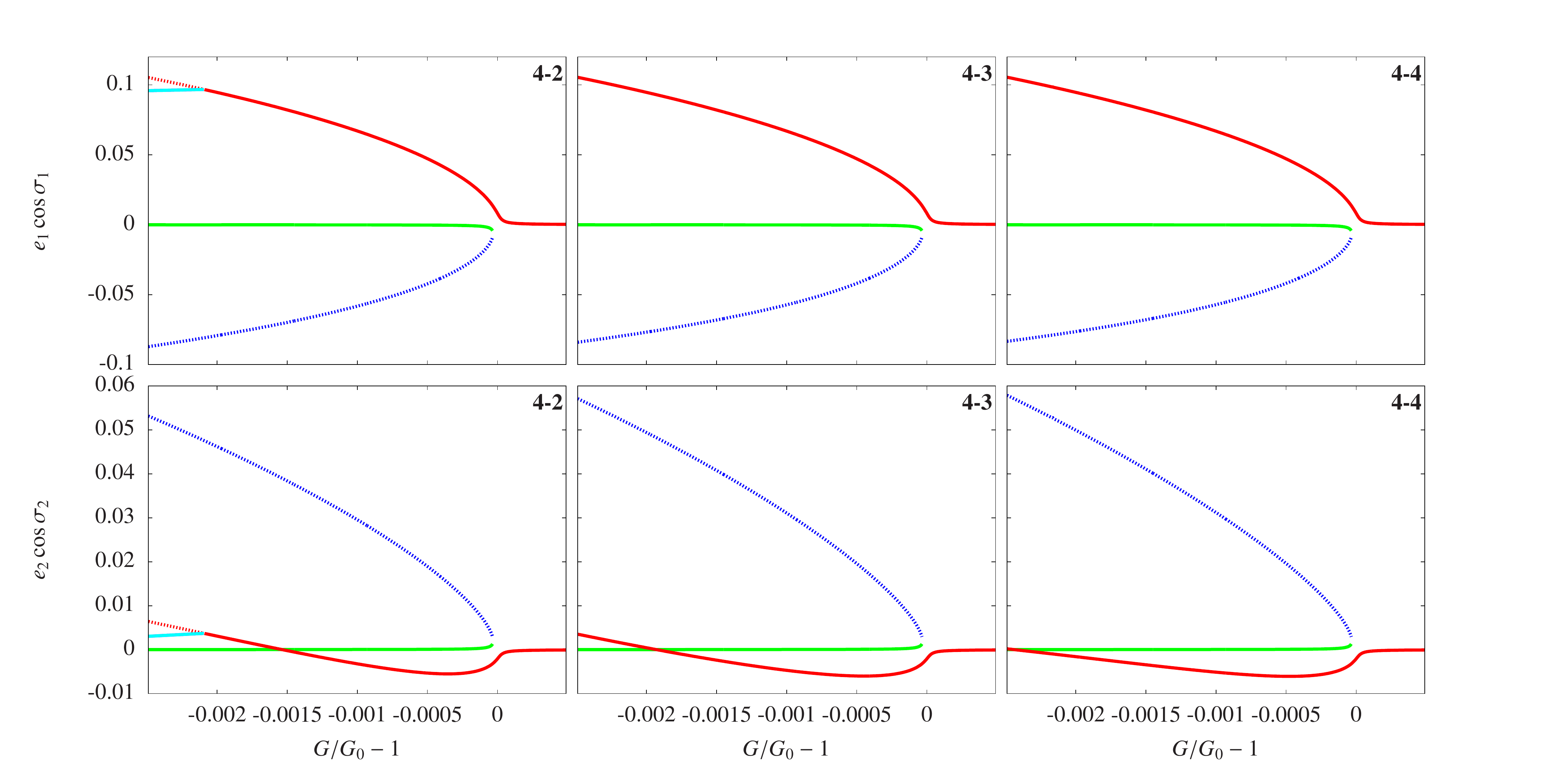}
    \caption{2:1 ACR positions in the directions of $\cos\sigma_i$ as functions of $G$.
    The conditions are the same as for Fig.~\ref{Figc} but
    the Hamiltonian is developed up to degree 4-2 (left), 4-3 (center), 4-4 (right).
    At degrees 4-3 and 4-4, all ACR have $\sin\sigma_1=\sin\sigma_2=0$.
    For the degree 4-2, the positions of ACR in the directions of $\sin\sigma_i$ are plotted in Fig.~\ref{Figf}.}
    \label{Fige}
  \end{figure*}
}

\newcommand\figf{
  \begin{figure}
    \centering
    \includegraphics[width=8cm,trim = 0.5cm 0cm 0cm 1cm, clip]{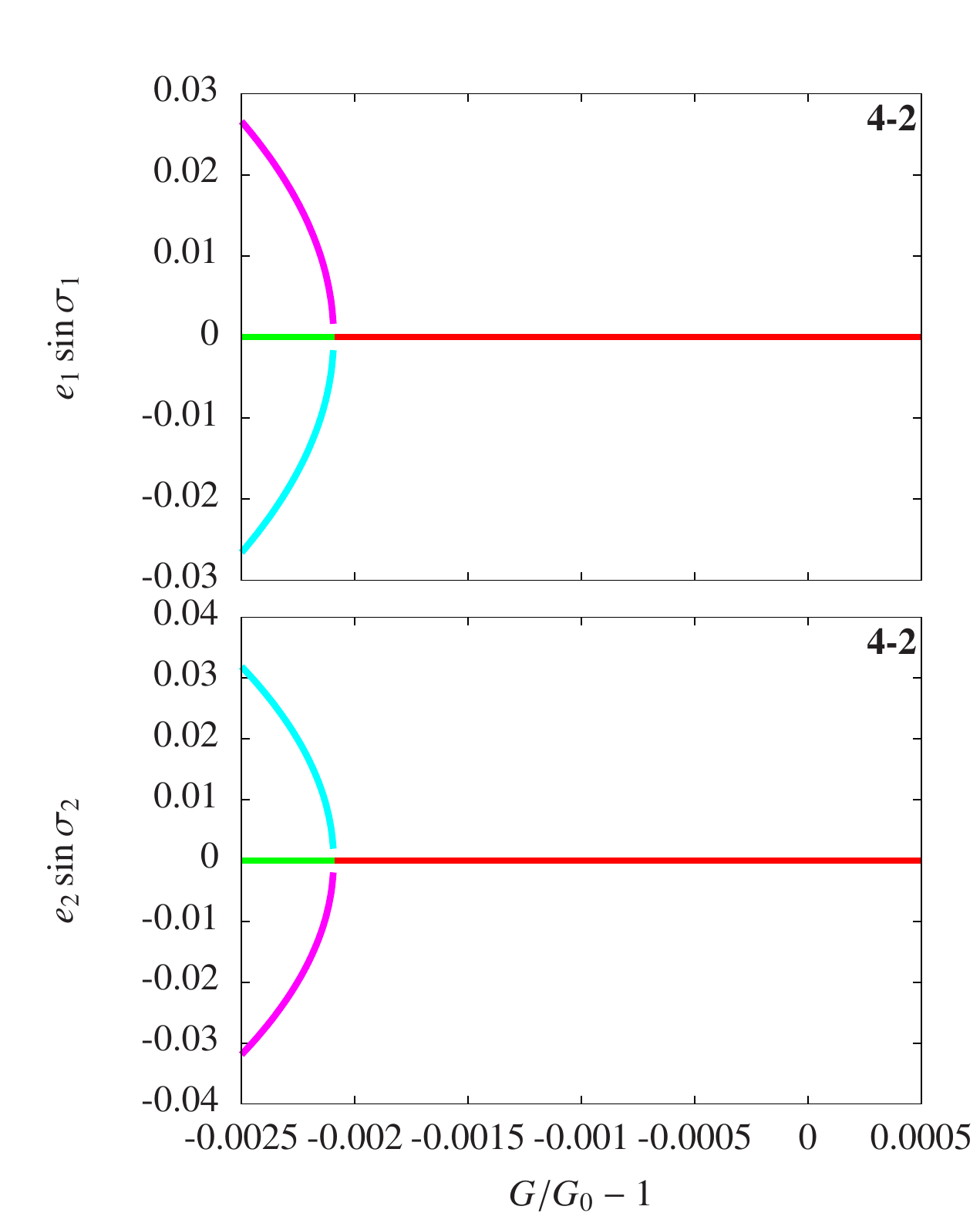}
    \caption{2:1 ACR positions in the directions of $\sin\sigma_i$ as functions of $G$.
    The conditions are the same as for Fig.~\ref{Figc} but
    the Hamiltonian is developed up to degree 4-2.}
    \label{Figf}
  \end{figure}
}

\newcommand\figg{
  \begin{figure}
    \centering
    \includegraphics[width=8cm,trim = 0.5cm 0cm 0cm 1cm, clip]{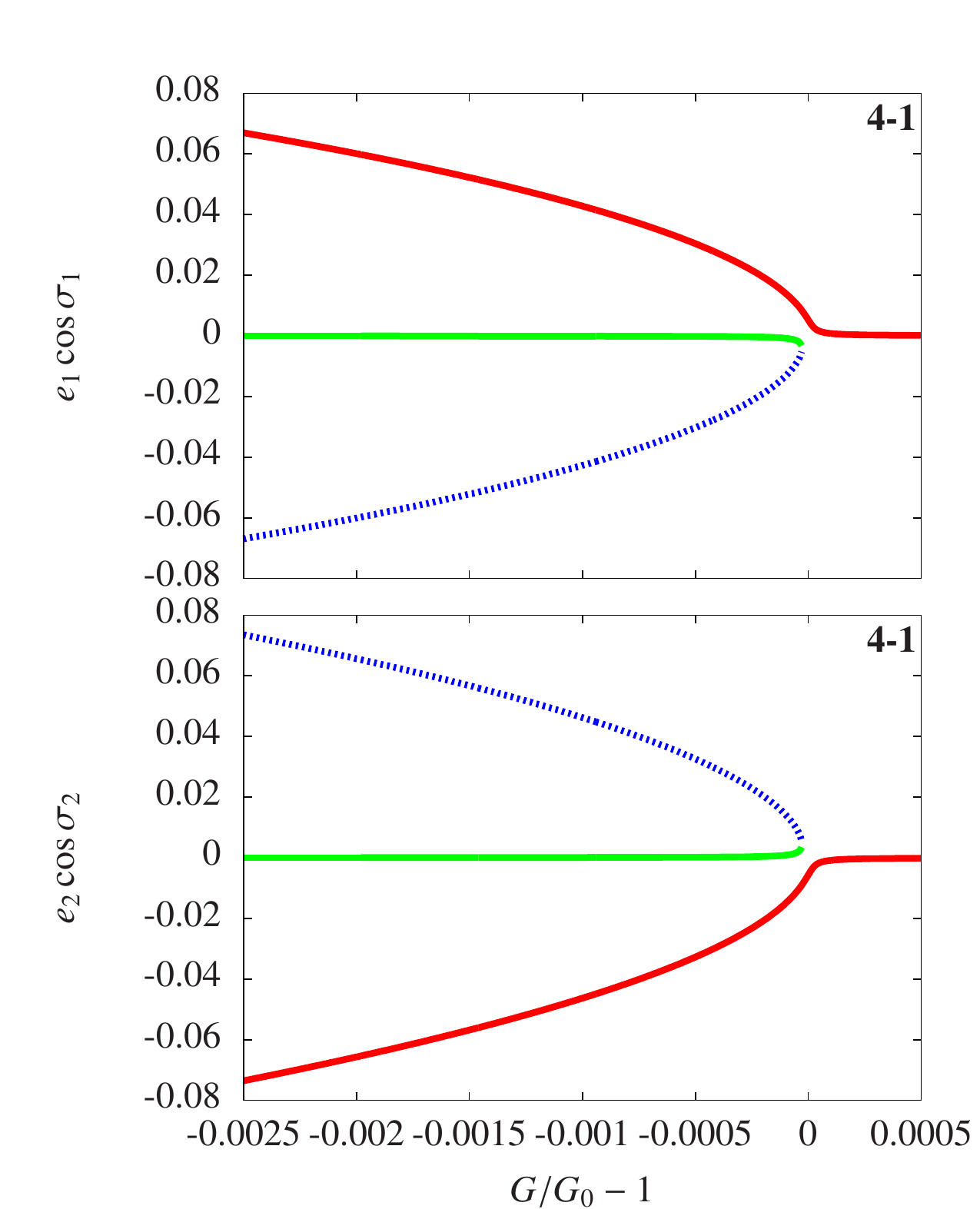}
    \caption{3:2 ACR positions as functions of $G$.
    The star mass is given by $m_0 = m_\sun$ and planets masses by $m_1=m_2=m_\earth$.
    The Hamiltonian is developed up to degree 4-1.
    Elliptic (stable) ACR are plotted using continuous lines whereas
    hyperbolic (unstable) ACR are plotted using dashed lines.
    At this degree of development, all ACR have $\sin\sigma_1=\sin\sigma_2=0$.}
    \label{Figg}
  \end{figure}
}

\newcommand\figh{
  \begin{figure*}
    \centering
    \includegraphics[width=15cm,trim = 1cm 0cm 4cm 2.5cm, clip]{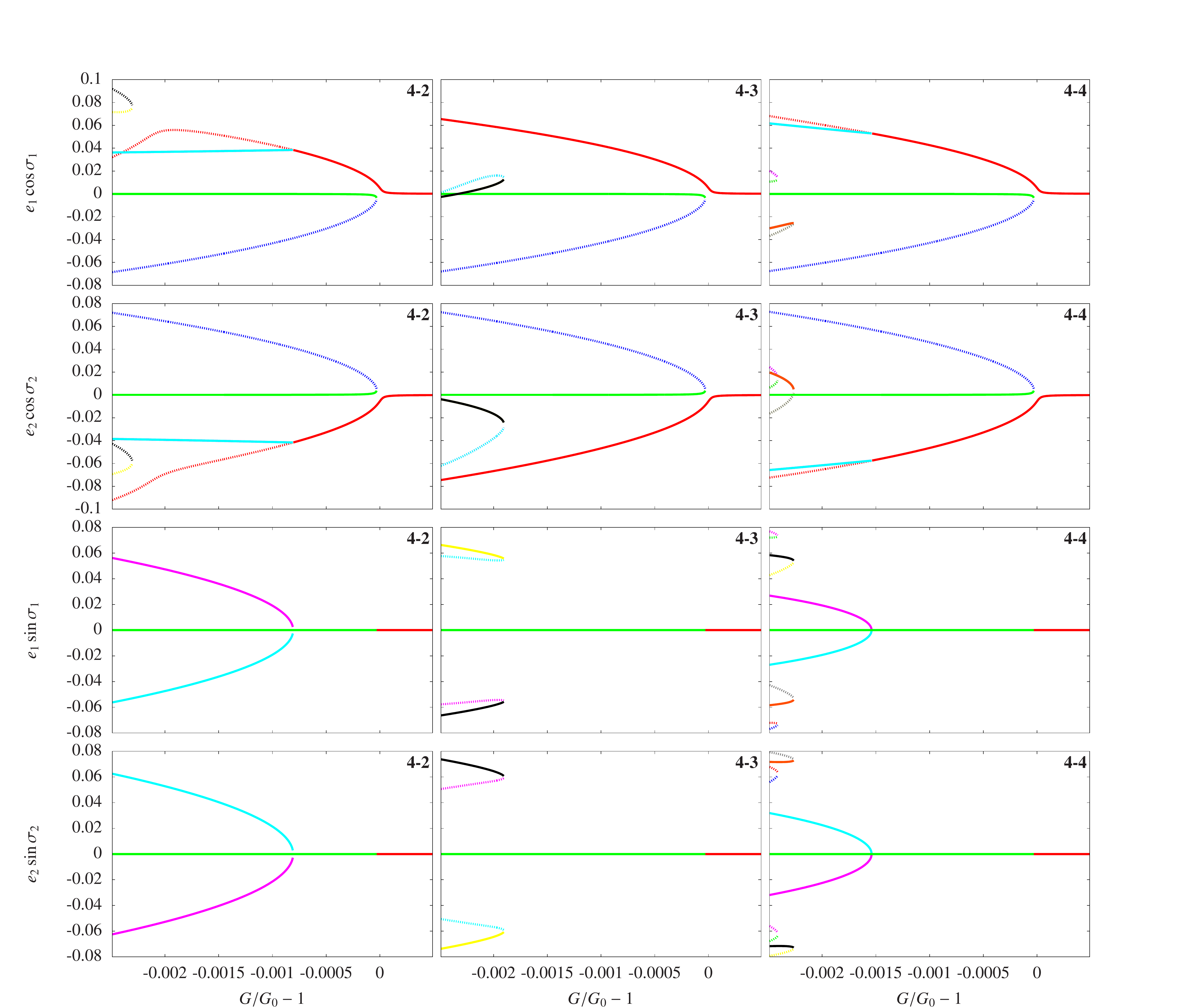}
    \caption{3:2 ACR positions as functions of $G$ in the same conditions as for Fig.~\ref{Figg} but
    the Hamiltonian is developed up to degree 4-2 (left), 4-3 (center), 4-4 (right).}
    \label{Figh}
  \end{figure*}
}

\newcommand\figi{
  \begin{figure}
    \centering
    \includegraphics[width=8cm,trim = 1cm 0cm 0cm 1cm, clip]{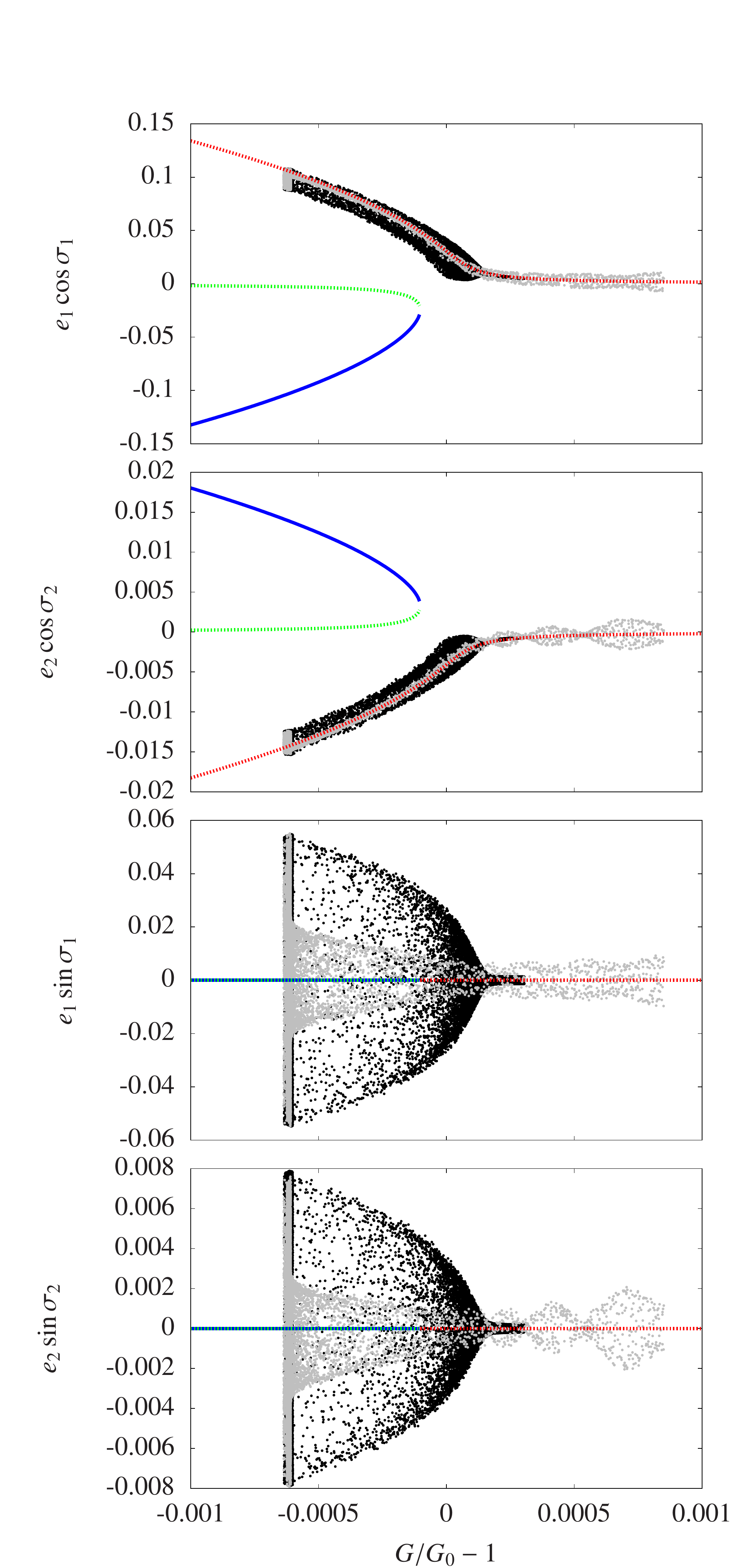}
    \caption{Superposition of the successive positions of the system in simulation $S1$ over
    the 3:2 ACR positions as functions of $G$.
    ACR positions are computed using our model with the parameters of the considered system (masses).
    The colors of ACR are consistent with graphs of section~\ref{sec:conservative} but
    we inverted continuous and dashed lines in order to improve the visibility of the simulation dots.
    The gray dots correspond to the migration phase and go from the right to the left of the graph
    whereas the black ones correspond to the tidal circularization phase and go from the left to the right.}
    \label{Figi}
  \end{figure}
}

\newcommand\figj{
  \begin{figure}
    \centering
    \includegraphics[width=8cm,trim = 0cm 0cm 1cm 0cm, clip]{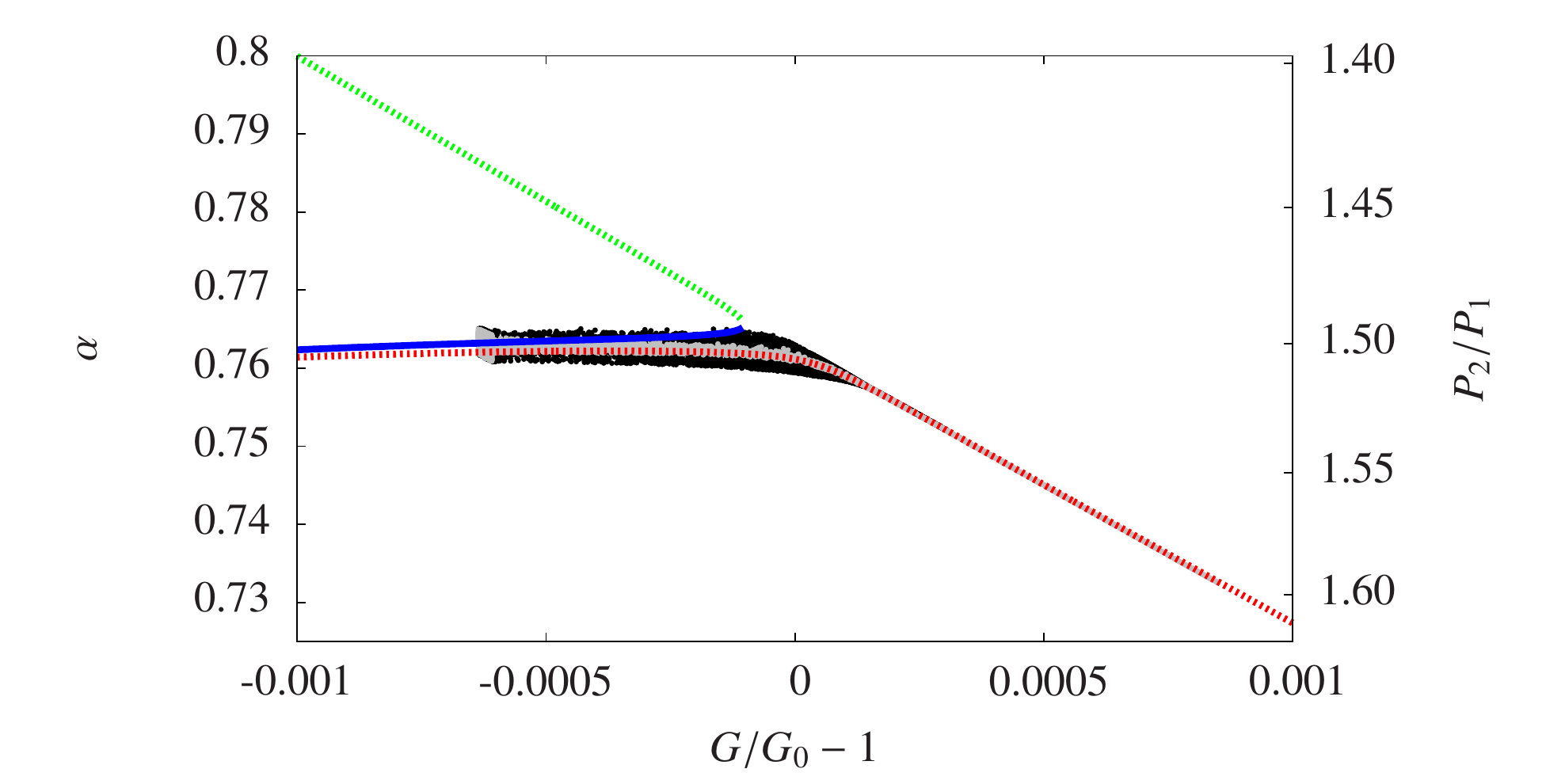}
    \caption{Superposition of the successive positions of the system in simulation $S1$ over
    the 3:2 ACR positions in the plane $(G,\alpha)$.
    ACR positions are computed using our model with the parameters of the considered system (masses).
    The colors of ACR are consistent with graphs of section~\ref{sec:conservative} but
    we inverted continuous and dashed lines in order to improve the visibility of the simulation dots.
    The gray dots correspond to the migration phase and go from the right to the left of the graph
    whereas the black ones correspond to the tidal circularization phase and go from the left to the right.}
    \label{Figj}
  \end{figure}
}

\newcommand\figk{
  \begin{figure}
    \centering
    \includegraphics[width=8cm,trim = 0.5cm 0cm 0cm 1cm, clip]{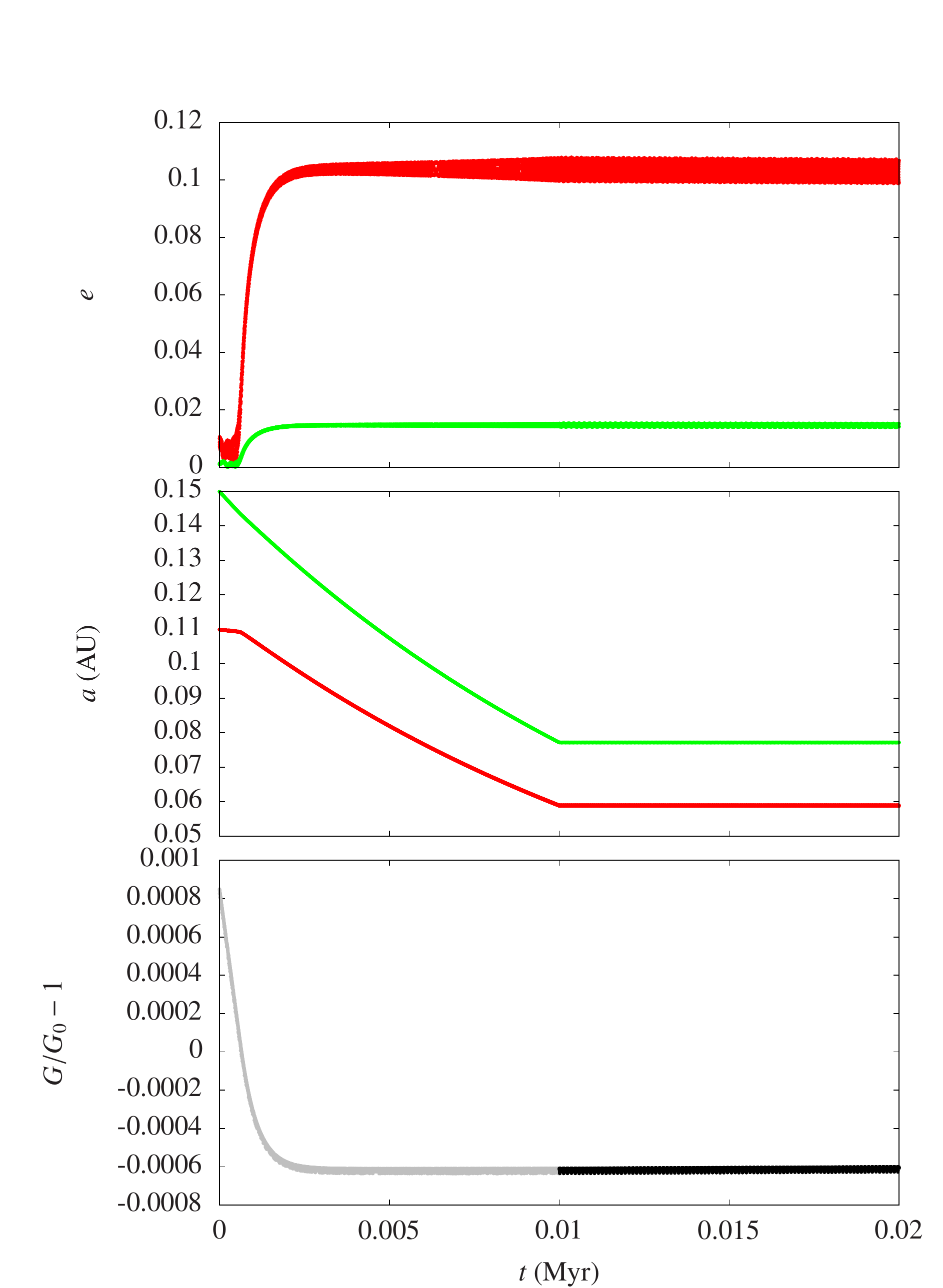}
    \caption{Evolution of eccentricities (top), semi-major axes (middle), and the parameter $G$ (bottom)
    during the first $20\ \mathrm{kyr}$ of simulation $S1$.
    The red curves correspond to planet 1.
    The green curves correspond to planet 2.
    The gray part of the $G$ curve (bottom) corresponds to the migration phase whereas
    the black part corresponds to the tidal circularization phase.}
    \label{Figk}
  \end{figure}
}

\newcommand\figl{
  \begin{figure}
    \centering
    \includegraphics[width=8cm,trim = 0.5cm 0cm 0.5cm 0cm, clip]{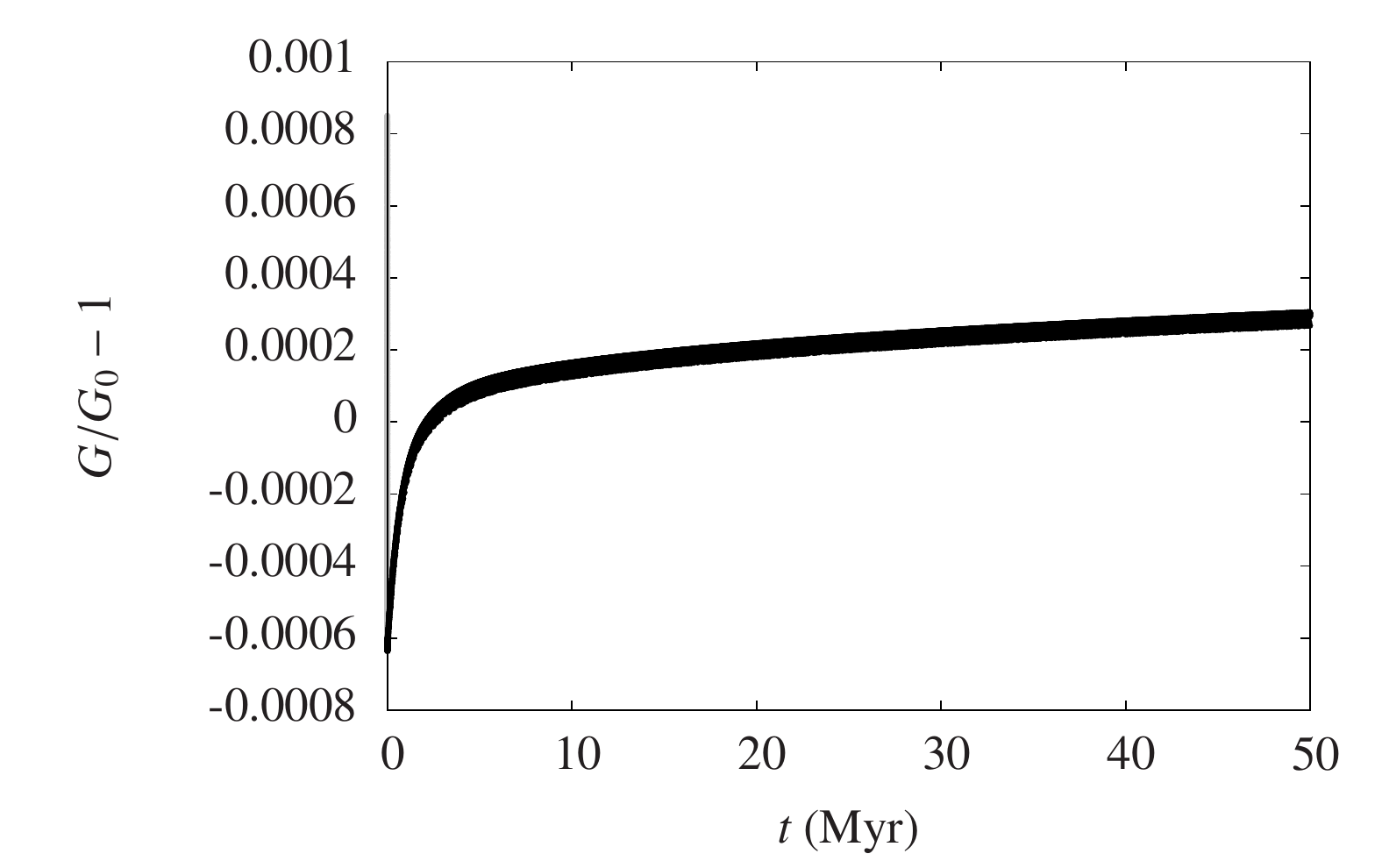}
    \caption{Long term evolution of the parameter $G$ during simulation $S1$.}
    \label{Figl}
  \end{figure}
}

\newcommand\figm{
  \begin{figure}
    \centering
    \includegraphics[width=8cm,trim = 0cm 0cm 0.5cm 0cm, clip]{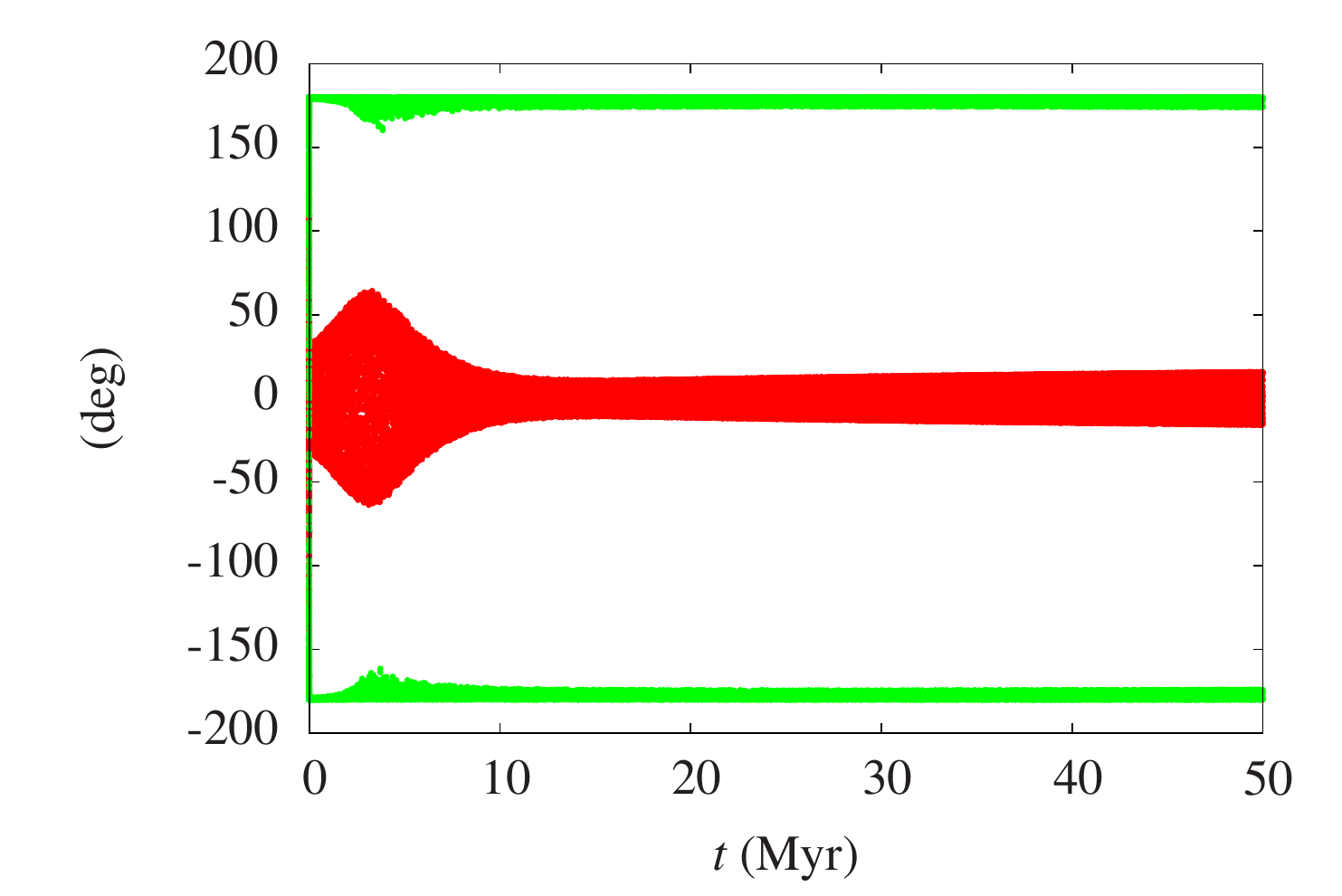}
    \caption{Evolution of the resonant angle $\sigma_1$ (red) and the difference of periastrons
    $\Delta \omega = \sigma_2 - \sigma_1$  (green) in simulation $S1$.}
    \label{Figm}
  \end{figure}
}

\newcommand\figii{
  \begin{figure}
    \centering
    \includegraphics[width=8cm,trim = 1cm 0cm 0cm 1cm, clip]{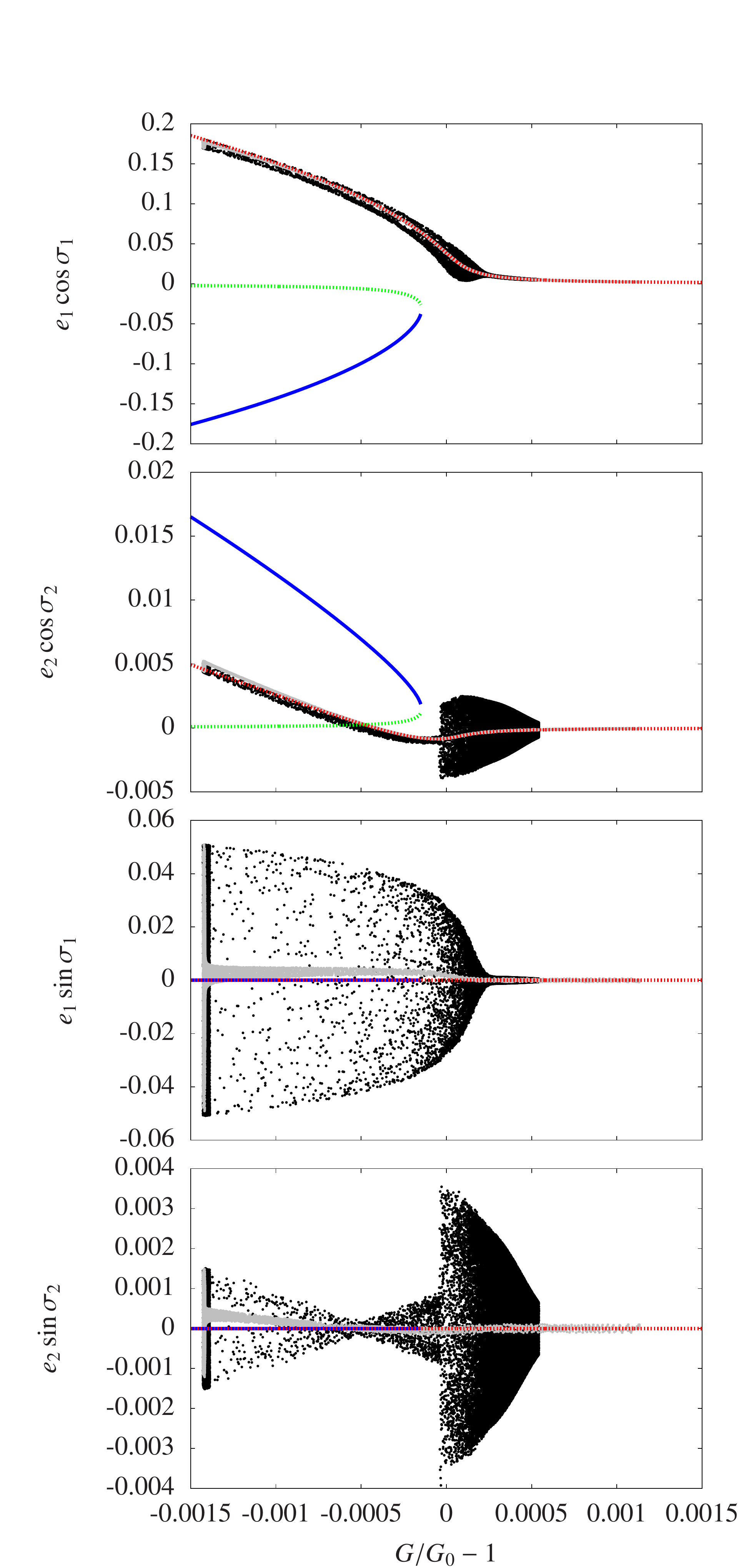}
    \caption{Superposition of the successive positions of the system in simulation $S2$ over
    the 2:1 ACR positions as functions of $G$.
    ACR positions are computed using our model with the parameters of the considered system (masses).
    The colors of ACR are consistent with graphs of section~\ref{sec:conservative} but
    we inverted continuous and dashed lines in order to improve the visibility of the simulation dots.
    The gray dots correspond to the migration phase and go from the right to the left of the graph
    whereas the black ones correspond to the tidal circularization phase and go from the left to the right.}
    \label{Figii}
  \end{figure}
}

\newcommand\figjj{
  \begin{figure}
    \centering
    \includegraphics[width=8cm,trim = 0cm 0cm 1cm 0cm, clip]{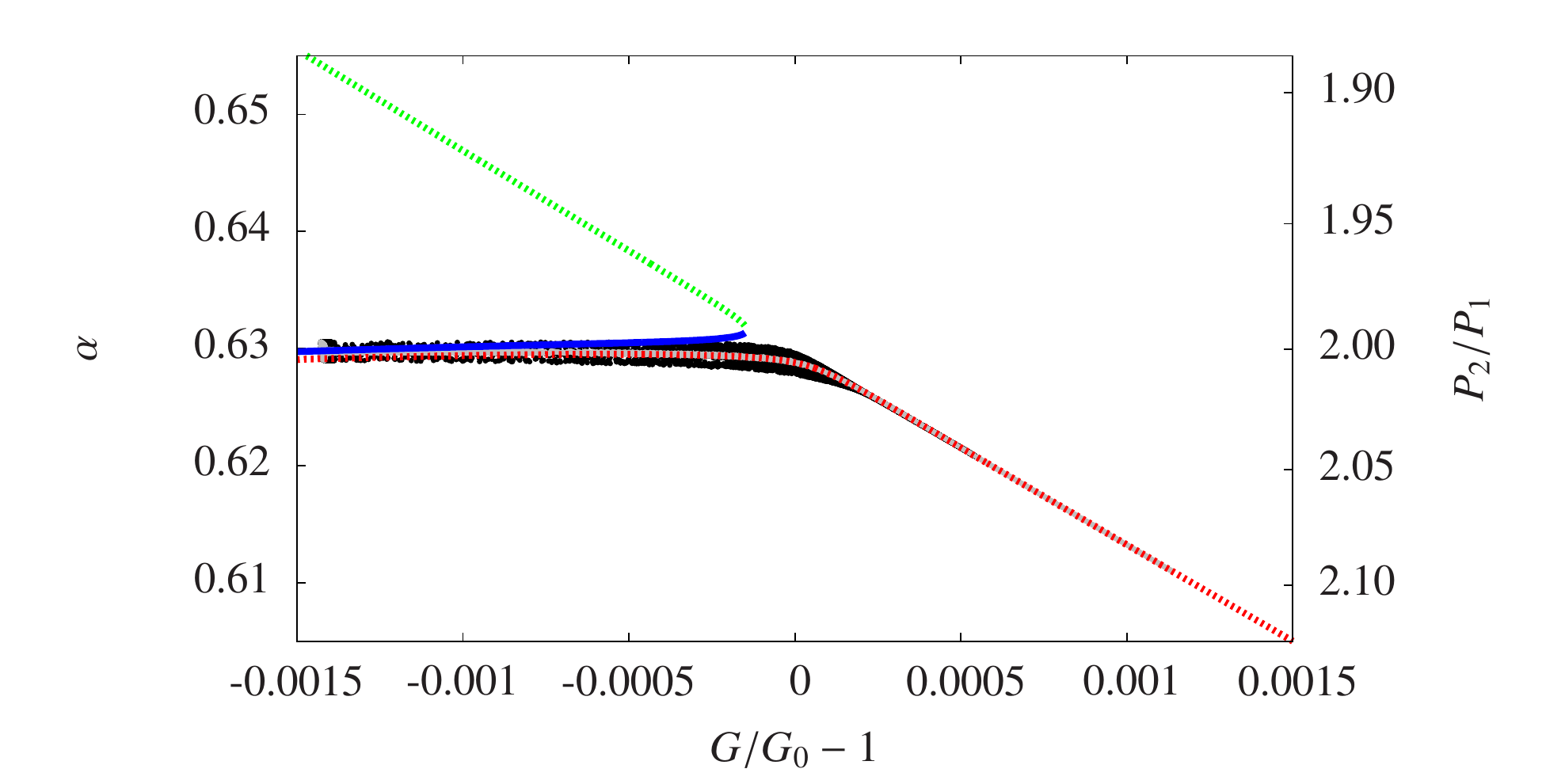}
    \caption{Superposition of the successive positions of the system in simulation $S2$ over
    the 2:1 ACR positions in the plane $(G,\alpha)$.
    ACR positions are computed using our model with the parameters of the considered system (masses).
    The colors of ACR are consistent with graphs of section~\ref{sec:conservative} but
    we inverted continuous and dashed lines in order to improve the visibility of the simulation dots.
    The gray dots correspond to the migration phase and go from the right to the left of the graph
    whereas the black ones correspond to the tidal circularization phase and go from the left to the right.}
    \label{Figjj}
  \end{figure}
}

\newcommand\figkk{
  \begin{figure}
    \centering
    \includegraphics[width=8cm,trim = 0.5cm 0cm 0cm 1cm, clip]{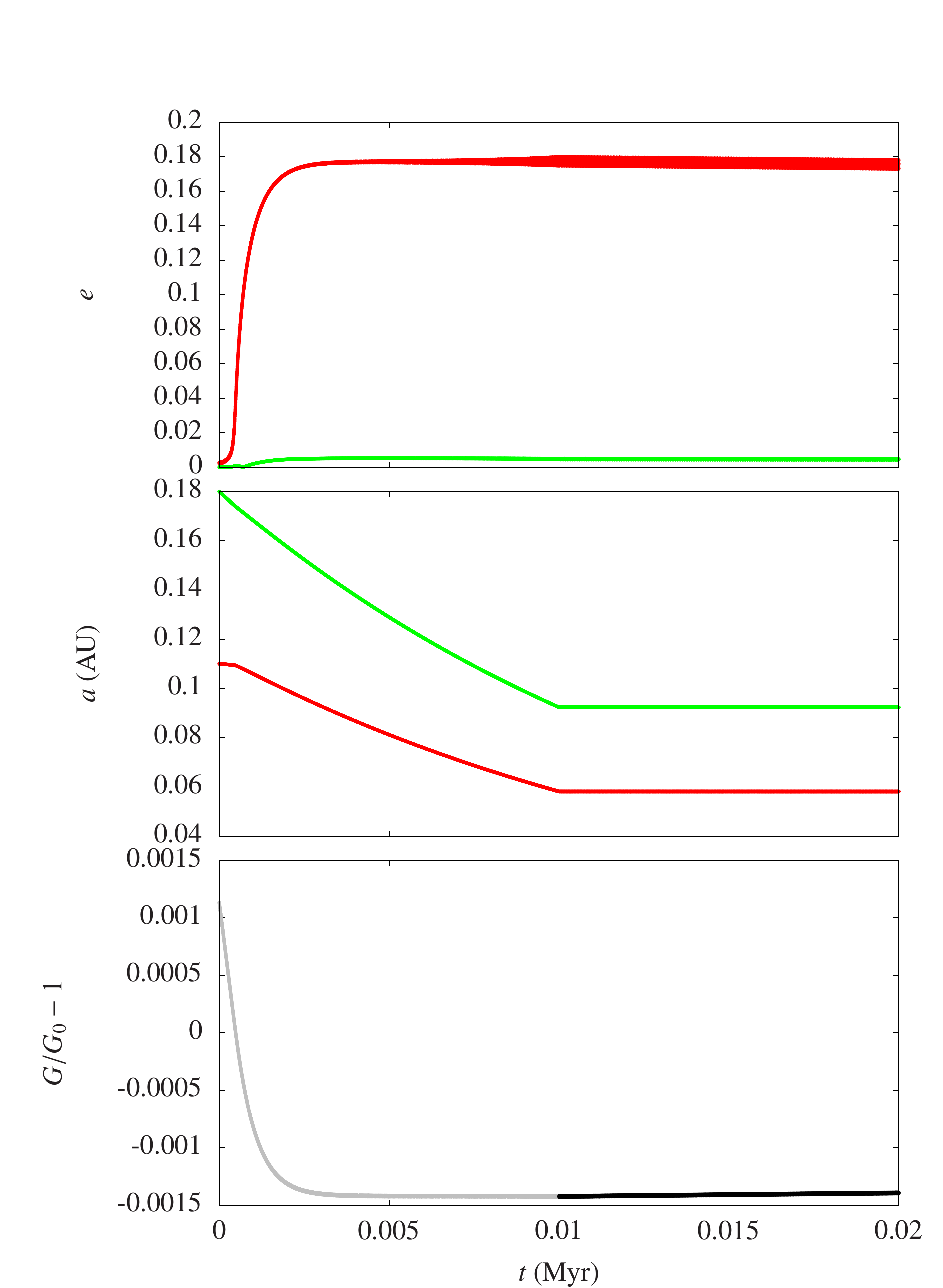}
    \caption{Evolution of the eccentricities (top) the semi-major axes (middle) and the parameter $G$ (bottom)
    during the first $20\ \mathrm{kyr}$ of simulation $S2$.
    The red curves correspond to planet 1.
    The green curves correspond to planet 2.
    The gray part of the $G$ curve (bottom) corresponds to the migration phase whereas
    the black part corresponds to the tidal circularization phase.}
    \label{Figkk}
  \end{figure}
}

\newcommand\figll{
  \begin{figure}
    \centering
    \includegraphics[width=8cm,trim = 0.5cm 0cm 0.5cm 0cm, clip]{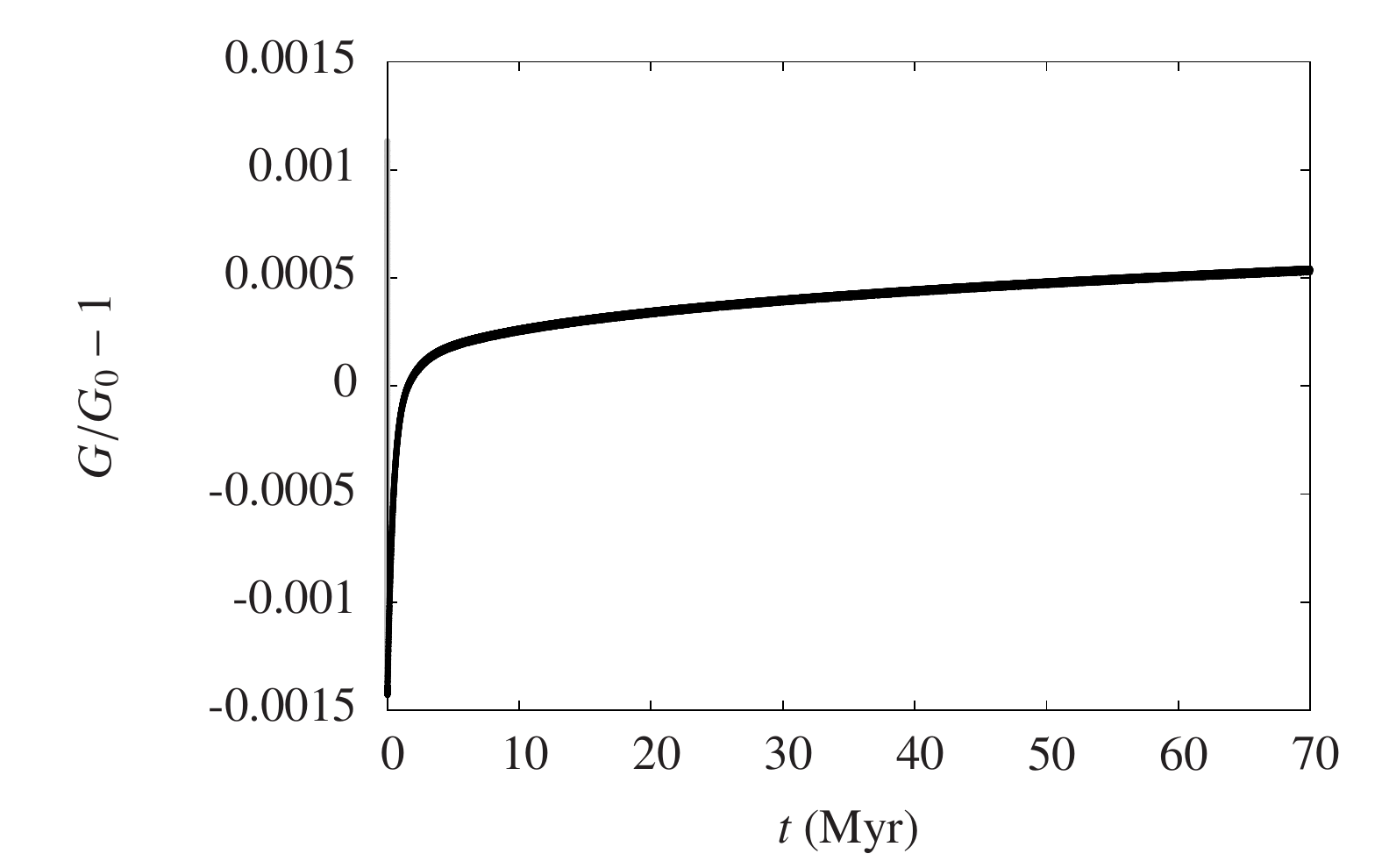}
    \caption{Long term evolution of the parameter $G$ during simulation $S2$.}
    \label{Figll}
  \end{figure}
}

\newcommand\figmm{
  \begin{figure}
    \centering
    \includegraphics[width=8cm,trim = 0cm 0cm 0.5cm 0cm, clip]{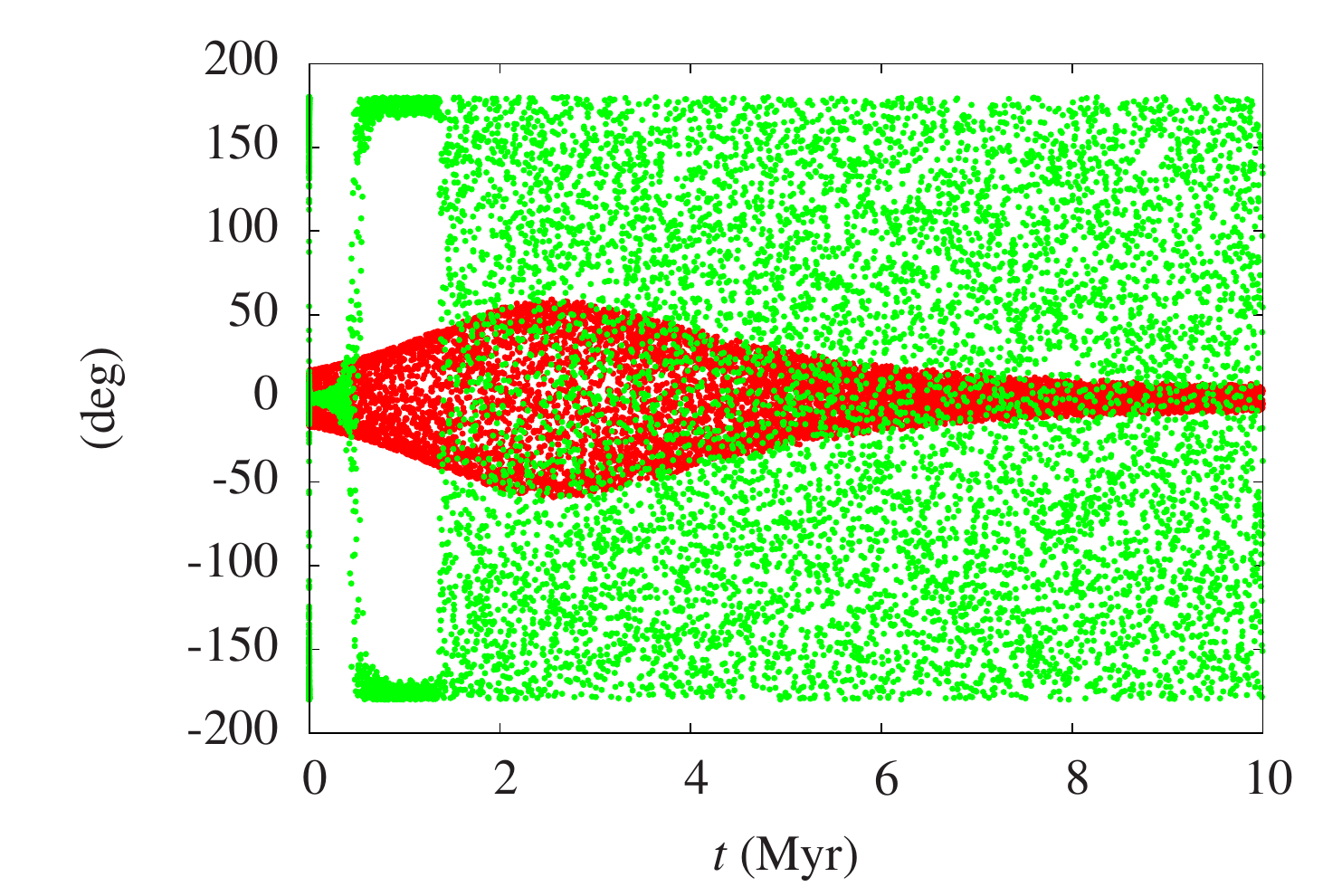}
    \caption{Evolution of the resonant angle $\sigma_1$ (red) and the difference of periastrons
    $\Delta \omega = \sigma_2 - \sigma_1$  (green) during the first $10\ \mathrm{Myr}$ of simulation $S2$.}
    \label{Figmm}
  \end{figure}
}

\newcommand\fign{
  \begin{figure}
    \centering
    \includegraphics[width=8cm]{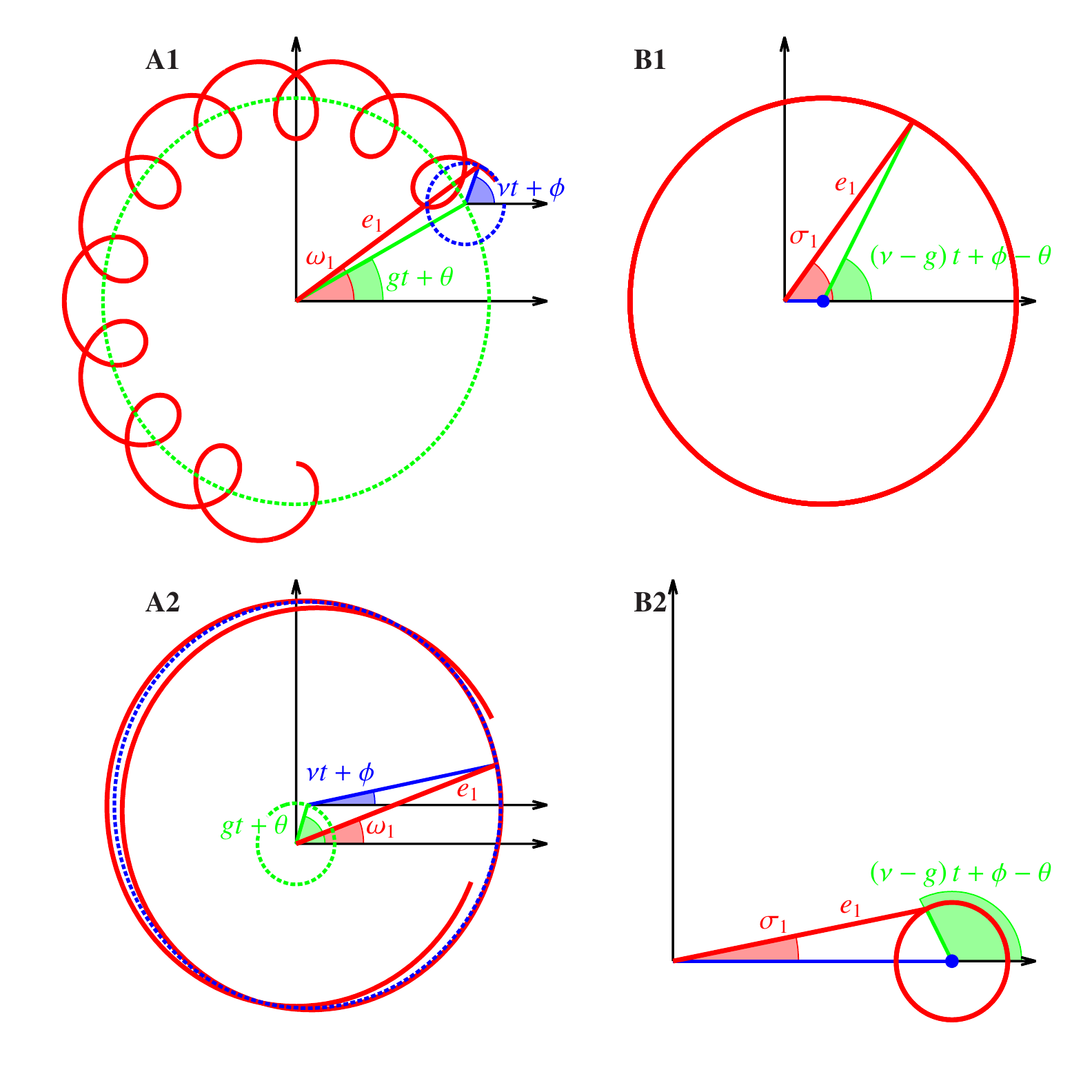}
    \caption{Illustration of the artificial libration of the resonant angles.
    We plot the evolution of the complex variables $e_1 \expo{\ImUnit \omega_1}$ (A1, A2) and
    $e_1 \expo{\ImUnit \sigma_1}$ (B1, B2) in the case of a strong secular eigenmode (A1, B1)
    and in the case of a severely damped one (A2, B2).
    For the sake of simplicity, we suppose that there is only one secular eigenmode (of frequency $g$),
    but the reasoning would be the same with more.
    We also suppose that there is only one high frequency term,
    corresponding to the nearest first order resonance,
    and whose frequency is $\nu = (p+1) n_2 - p n_1$.
    The secular contribution is plotted in green in all graphs whereas the high frequency term contribution
    is plotted in blue.
    The red curves correspond to the sums of both contributions.
    When there is a strong secular eigenmode,
    the high frequency term only brings small perturbations around the secular
    evolution of the argument of periastron $\omega_1$ (A1)
    and the resonant angle $\sigma_1$ (B1) which circulates.
    When the eigenmode is damped, the evolution of $\omega_1$ (A2)
    and $\sigma_1$ (B2) are dominated by the high frequency $\nu$.
    The amplitude corresponding to $\nu$ does not change between A1, B1 and A2, B2 (the scale changes) but
    in B2 $\sigma_1$ appears to librate
    because the circulation center (marked with a blue dot in B1, B2) is not  at zero.
    }
    \label{Fign}
  \end{figure}
}

\newcommand\figo{
  \begin{figure}[h]
    \centering
    \includegraphics[width=8cm]{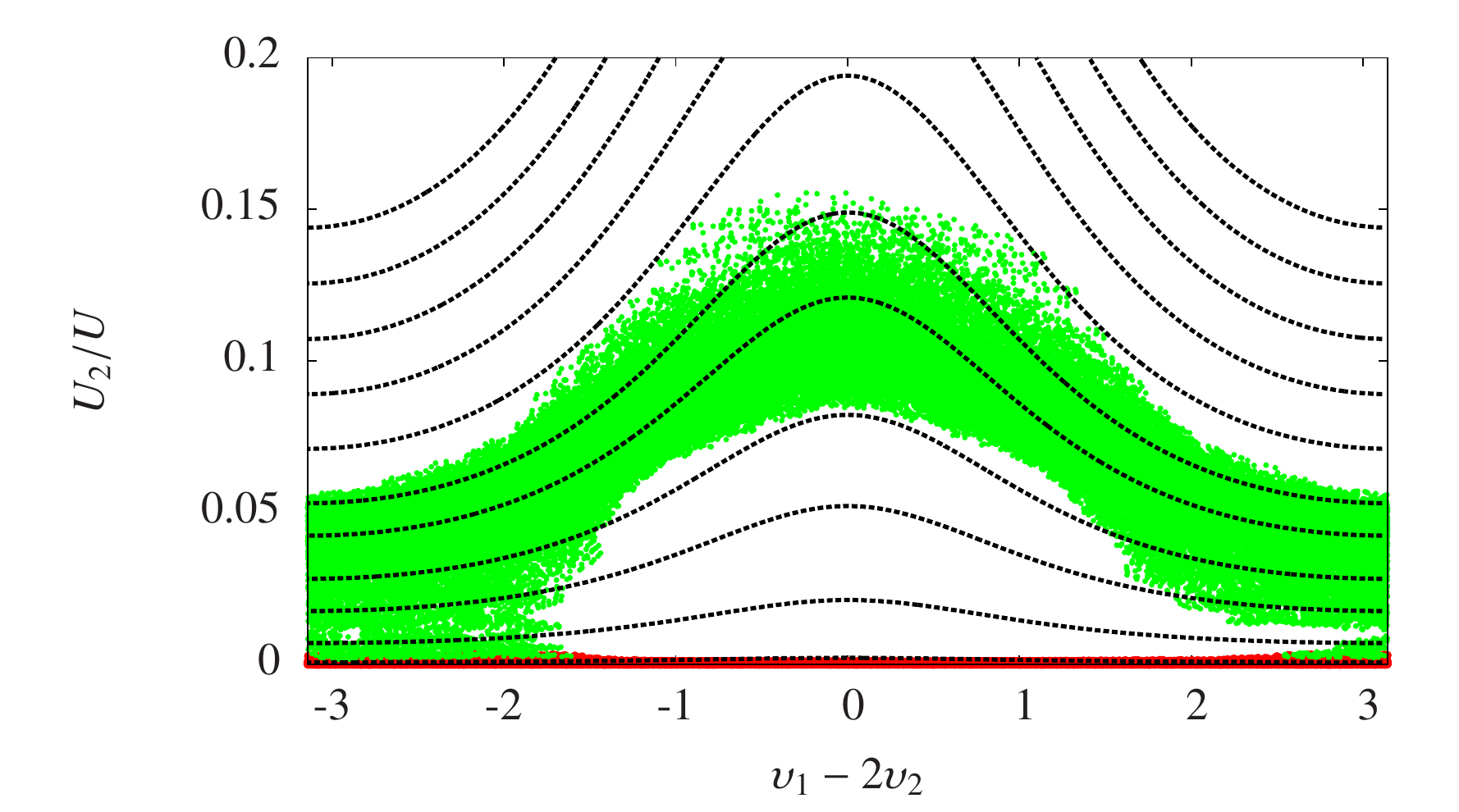}
    \caption{Superposition of the trajectory of the system $S2$ during the event (around $t=1.4\ \mathrm{Myr}$)
    with the energy levels curves of the 2:1 resonance between both proper modes around the ACR solution.
    The successive positions of the system just before the event are marked with red dots while the positions
    of the system just after the event are marked with green dots.
    }
    \label{Figo}
  \end{figure}
}

\begin{document}

\title{Dissipation in planar resonant planetary systems}
\author{J.-B. Delisle\inst{1}
\and J. Laskar\inst{1}
\and A. C. M. Correia\inst{1,2}
\and G. Bou\'e\inst{1,3}
}

\institute{Astronomie et Syst\`emes Dynamiques, IMCCE-CNRS UMR8028, Observatoire de Paris, UPMC,
77 Av. Denfert-Rochereau, 75014~Paris, France\\
\email{delisle@imcce.fr}
\and Department of Physics, I3N, University of Aveiro, Campus Universit\'ario de Santiago,
3810-193 Aveiro, Portugal
\and Centro de Astrof\'isica, Universidade do Porto, Rua das Estrelas, 4150-762 Porto, Portugal
}

\date{\today}

\abstract{
Close-in planetary systems detected by the Kepler mission present an excess of
periods ratio that are just slightly larger than some low order resonant values.
This feature occurs naturally when resonant couples undergo
dissipation that damps the eccentricities.
However, the resonant angles appear to librate at the end of the
migration process, which is often believed to be an evidence that the
systems remain in resonance.

Here we provide an analytical model for the dissipation in resonant
planetary systems valid for low eccentricities.
We confirm that dissipation accounts for an excess of pairs that lie
just aside from the nominal periods ratios, as observed by the Kepler mission.
In addition, by a global analysis of the phase space of the problem,
we  demonstrate that these final pairs are
non-resonant. Indeed, the separatrices that exist in the resonant systems disappear
with the dissipation, and remains only a circulation of the orbits
around a single elliptical fixed point.
Furthermore,  the apparent libration of the resonant angles can be explained using the
classical secular averaging method. We show that this artifact
is only due to the severe damping of
the amplitudes of the eigenmodes in the secular motion.
}


\maketitle

\section{Introduction}
Dissipation due to tidal interactions is a possible mechanism that explains the
abundance of planetary systems that lie near but not at exact mean-motion commensurability.
\citet{papaloizou_dynamics_2010} (in the case of three planets Laplace resonances) and \citet{papaloizou_tidal_2011} (for two planets resonances) showed that planets that have been temporarily locked in resonance due to differential migration
could have their periods ratios to depart from strict commensurability due to the circularization of their orbits
by tidal interactions with the star.
More recently, \citet{batygin_dissipative_2012} and \citet{lithwick_resonant_2012}
use similar effects to explain the excess of systems of two planets that lie near resonances but with planets slightly
farther from each other than the nominal mean-motion commensurability ratio.

One of the most intriguing features that is common in these different studies is the observation
that resonant angles continue to librate far from exact commensurability and the authors leave unanswered the question of determining if these systems are
in resonance or not.
It seems important to clarify this point and to understand why resonant angles can librate so far from exact commensurability.

Our analysis is based on the study of the phase space of two planets in mean-motion resonance (MMR) in the conservative case
(without any dissipation). This is the object of section \ref{sec:conservative}.
We pay a particular attention to apsidal corotation resonances (ACR) which play a major role in the dynamics of these systems
and in the understanding of the topology of the phase space.
ACR have been extensively studied both in the asteroidal restricted problem \citep[e.g.][]{ferraz-mello_symmetrical_1993} and the planetary problem \citep[e.g.][]{hadjidemetriou_2002,michtchenko_stationary_2006}.

Most of the studies on the subject have been made using numerical (or semi-analytical) models that remain
valid for arbitrary values of the eccentricities but which do not always provide a global picture of the dynamics.
In the present study we are only concerned in the dynamics at low eccentricities since our aim is to understand the motion
at the end of the circularization process.
A completely analytical model is thus well suited in this case.
Analytical studies of planetary MMR have already be done up to degree two in eccentricities
in the cases of the 2:1 \citep{callegari_dynamics_2004} and the 3:2 \citep{callegari_dynamics_2006} MMR.

The dissipative case (studied in section \ref{sec:dissipation}) is modeled using the conservative case as a basis and
very simple and general prescriptions for the dissipation.
Our study is mainly aimed at understanding the impact of tides on the dynamics of resonant planets pairs and,
in this case, we follow the prescriptions introduced by \citet{papaloizou_tidal_2011}.
However we show that the differential migration process that allows a resonant locking of both planets can also be accounted
for in our model. This process has already been widely studied
\citep[e.g.][]{lee_dynamics_2002, ferraz-mello_evolution_2003,lee_diversity_2004,beauge_planetary_2006},
and is also considered in \citet{papaloizou_dynamics_2010,papaloizou_tidal_2011}.
Eventually, we treat this perturbation of the conservative case by following the lines of
\citet{laskar_tidal_2012}.

In section \ref{sec:secular} we show that the final state of the resonant systems that undergo a circularization process
is very well characterized by the secular normal form. We explain, with the secular problem, why resonant angles
appear to librate even far from resonances.

Finally, we present in section \ref{sec:simulation} the results of two
numerical simulations that confirm and illustrate the different
mechanisms that we highlight with our analytical model.

\section{Dynamics of two resonant planets in the conservative case}\label{sec:conservative}
\subsection{Model}
We study in this section the case of two planets orbiting a star in the same plane
and without any dissipative force. This problem is usually referred to as the planar planetary three body problem.
We note with a subscript $1$ the internal planet and $2$ the external one.
The star is referred to as body $0$.
Masses are noted $m_i$.
For both planets we define:
$\mu_i = \mathcal{G} (m_0 + m_i)$,
and $\beta_i = m_0 m_i/(m_0 + m_i)$,
where $\mathcal{G}$ is the gravitational constant.
We note $a_i$ the semi-major axis of planet $i$, and
$e_i$ its eccentricity.

We use Delaunay canonical pairs of astrocentric coordinates for both planets.
The actions are the circular angular momentum and the angular momentum of both planets:
\begin{eqnarray}
  \scaled{\Lambda}_i &=& \beta_i \sqrt{\mu_i a_i},\\
  \scaled{G}_i &=& \scaled{\Lambda}_i \sqrt{1-e_i^2} = \beta_i \sqrt{\mu_i a_i (1-e_i^2)},
\end{eqnarray}
The associated angles are the mean anomalies ($M_i$) and the arguments of periastron ($\omega_i$) of both planets.

The Hamiltonian of the system reads, in these coordinates:
\begin{equation}
  \scaled{\mathcal{H}} = \scaled{\mathcal{H}}_0 (\scaled{\Lambda})
    + \scaled{\mathcal{H}}_1 (\scaled{\Lambda}, \scaled{G}, M, \omega),
\end{equation}
where $\scaled{\mathcal{H}}_0$ is the Keplerian part
and $\scaled{\mathcal{H}}_1$ is the perturbative part due to planet-planet interactions
taking into account both direct and indirect effects.
The Keplerian part depends only on $\scaled{\Lambda}_i$ ($i=1,2$):
\begin{equation} \label{eq:H0}
  \scaled{\mathcal{H}}_0 = - \sum_{i=1}^2 \frac{\mu_i^2 \beta_i^3}{2 \scaled{\Lambda}_i^2}.
\end{equation}
Whereas the perturbative part depends on all eight Delaunay coordinates. We do not need to express the explicit form of $\scaled{\mathcal{H}}_1$ at this point but it could be seen as a Fourier series of all four angles (with coefficients depending on actions).

At first sight, we have to deal with a four degrees of freedom differential system.
However, we will now introduce two well-known transformations which will reduce
this problem to two degrees of freedom.
The first reduction makes use of the total angular momentum conservation and is always valid.
The second one corresponds to an averaging of the equations of motion and
is only valid near a specified mean-motion resonance (MMR).
More precisely, it is only valid far from all other MMR.

\subsubsection{Conservation of the total angular momentum}
In the Delaunay coordinates system, the total angular momentum conservation
($\scaled{G} = \scaled{G}_1 + \scaled{G}_2$) is equivalent to the fact that
the Hamiltonian depends only on the value of the difference of arguments of periastron: $\Delta \omega=\omega_1-\omega_2$
and not on individual values of both angles.
The reduction \modif{resulting from the conservation} of the angular momentum is performed
by using the coordinates $\Delta \omega$, $\omega_2$
instead of $\omega_1$, $\omega_2$.
These two new angles are respectively canonically conjugated to the actions $\scaled{G}_1$, $\scaled{G}$.
Therefore, the system of coordinates
$(M_1,\scaled{\Lambda}_1,M_2,\scaled{\Lambda}_2,\Delta \omega,\scaled{G}_1, \omega_2, \scaled{G})$
is canonical and with these coordinates,
$\omega_2$ does not appear \modif{anymore} in the \modif{expression of the} Hamiltonian.
$\scaled{G}$ is a first integral of the equations of motion.
We can thus study the three degrees of freedom system constituted of
$(M_1,\scaled{\Lambda}_1,M_2,\scaled{\Lambda}_2,\Delta \omega,\scaled{G}_1)$
and consider $\scaled{G}$ as a parameter of this system.

\subsubsection{Averaging the Hamiltonian near a MMR}\label{sec:averaging}
We now suppose that the system is near a mean-motion commensurability of the form $(p+q)$:$p$, that is:
\begin{equation}
  - p \dot{M}_1 + (p+q) \dot{M}_2 \approx 0.
\end{equation}
We may then introduce the argument of the $(p+q)$:$p$ MMR:
\begin{equation}
  \sigma = - \frac{p}{q} M_1 + \left(1 + \frac{p}{q}\right) M_2,
\end{equation}
such as $\dot{\sigma} \approx 0\ $.

We then construct the resonant normal form to the first order of the planets masses (with respect to the stellar mass) by averaging over all rapid angles, i.e. all combinations of the mean anomalies that are not harmonics of $\sigma$.
By doing this averaging we need to introduce a new set of variables which corresponds to the averaged problem.
The change of coordinates is close to identity so we can identify both sets of coordinates in first approximation.
Strictly speaking the transformation to the initial set of coordinates reintroduces high frequencies
\citep[e.g.][]{morbidelli_modern_2002}.
Before constructing the normal form, we introduce the change of variables:
\begin{eqnarray}
  \scaled{I} &=& -\frac{q}{p} \scaled{\Lambda}_1,\\
  \Gamma &=& \left(1 + \frac{q}{p}\right) \scaled{\Lambda}_1 + \scaled{\Lambda}_2,
\end{eqnarray}
and we consider the canonical set of coordinates: $(\sigma,\scaled{I},M_2,\Gamma,\Delta\omega,\scaled{G}_1)$.
With this set of coordinates, the resonant normal form is obtained by averaging over $M_2$.
Thus, by definition, $M_2$ does not appear anymore in the averaged Hamiltonian,
and $\Gamma$ is a first integral of the averaged motion.
Therefore, the system is reduced to a two degrees of freedom problem
with four canonical coordinates: $(\sigma,\scaled{I},\Delta \omega,\scaled{G}_1)$
and two parameters: $\scaled{G}$ and $\Gamma$.
However this set of coordinates is not very well suited
for the study of low-eccentricities systems and
it is a lot more convenient to introduce rectangular canonically conjugated variables
similar to Poincar\'e variables.

\subsubsection{Rectangular coordinates}
We first introduce the two resonant angles which correspond to both planets
\citep[e.g.][]{ferraz-mello_symmetrical_1993}:
\begin{eqnarray}
  \sigma_1 &=& \sigma - \left(1+\frac{p}{q}\right) \Delta \omega
    = -\frac{p}{q} \lambda_1 + \left(1 + \frac{p}{q}\right) \lambda_2 - \omega_1,\\
  \sigma_2 &=& \sigma - \frac{p}{q} \Delta \omega
    = -\frac{p}{q} \lambda_1 + \left(1 + \frac{p}{q}\right) \lambda_2 - \omega_2,
\end{eqnarray}
where $\lambda_i = M_i + \omega_i$ is the mean longitude of planet $i$.
We complete this change of variables with the canonically conjugated actions:
\begin{eqnarray}
  \scaled{I}_1 &=& -\frac{p}{q} \scaled{I} - \scaled{G}_1
    = \scaled{\Lambda}_1 - \scaled{G}_1,\\
  \scaled{I}_2 &=& \left(1+\frac{p}{q}\right) \scaled{I} + \scaled{G}_1
    = \scaled{\Lambda}_2 - \scaled{G}_2 + \scaled{G} - \Gamma.
\end{eqnarray}
Note that, since $\scaled{G}$ and $\Gamma$ are constants of the motion, we can redefine the action $\scaled{I}_2$ as:
\begin{equation}
  \scaled{I}_2 = \scaled{\Lambda}_2 - \scaled{G}_2,
\end{equation}
without any change for $\scaled{I}_1$, $\sigma_1$ and $\sigma_2$.

The associated rectangular coordinates are then:
\begin{equation}\label{eq:rectcoord}
  \scaled{x}_i = \sqrt{\scaled{I}_i} \expo{\ImUnit \sigma_i}.
\end{equation}

\subsubsection{Elimination of the first integral $\Gamma$}
The dynamics of our two degrees of freedom system should depend on the value of both first integrals:
$\Gamma$ and $\scaled{G}$. However, it is possible to eliminate the dependency in $\Gamma$ by dividing all
actions by its value:
\begin{eqnarray}
  \Lambda_i &=& \frac{\scaled{\Lambda}_i}{\Gamma},\\
  G_i &=& \frac{\scaled{G}_i}{\Gamma},\\
  G &=& \frac{\scaled{G}}{\Gamma} = G_1 + G_2,\\
  I_1 &=& \frac{\scaled{I}_1}{\Gamma} = \Lambda_1 - G_1,\\
  I_2 &=& \frac{\scaled{I}_2}{\Gamma} = \Lambda_2 - G_2,\\
  x_i &=& \frac{\scaled{x}_i}{\sqrt{\Gamma}}.
\end{eqnarray}
With these renormalized variables, Hamilton equations are still valid if the Hamiltonian is also divided by $\Gamma$:
\begin{equation}
  \tempscaled{\mathcal{H}} = \frac{\scaled{\mathcal{H}}}{\Gamma}.
\end{equation}
However, the Hamiltonian $\tempscaled{\mathcal{H}}$, and thus the dynamics, still depends on the value of $\Gamma$.
The complete elimination of $\Gamma$ is achieved by renormalizing both energy and time scales:
\begin{eqnarray}
  \mathcal{H} &=& \Gamma^3 \tempscaled{\mathcal{H}} = \Gamma^2 \scaled{\mathcal{H}},\\
  \tau &=& \frac{t}{\Gamma^3}.
\end{eqnarray}
It can be verified that $\mathcal{H}$ does not depend anymore on $\Gamma$ and Hamilton's equations now reads:
\begin{equation}\label{eq:Hameq}
  \deriv{x_i}{\tau} = -\ImUnit \dpart{\mathcal{H}}{\overline{x}_i}.
\end{equation}
This means that the value of $\Gamma$ does not influence
the dynamics of the system except by changing the scales of distance, time and energy
(respectively by a factor of $\Gamma^2$, $\Gamma^3$ and $1/\Gamma^2$).
With this renormalization, we reduced the number of parameters of our system from two
($\scaled{G}$, $\Gamma$) to one ($G = \scaled{G}/\Gamma$).

\subsubsection{Explicit form of the Hamiltonian}
We need to express the Hamiltonian $\mathcal{H}$ as a function of $x_i$ ($i=1,2$) and $G$.
Let us start with the Keplerian part. The renormalized Keplerian part ($\mathcal{H}_0$)
has the same form as $\scaled{\mathcal{H}}_0$ (see \refeq{eq:H0}) if we replace $\scaled{\Lambda}_i$ with $\Lambda_i$.
Moreover, $\Lambda_i$ are simple functions of $x_i$ and $G$:
\begin{eqnarray}
    \Lambda_1 &=& - \frac{p}{q} \left(G + \mathcal{D} - 1 \right),\\
    \Lambda_2 &=& 1 - \left(1+\frac{q}{p}\right) \Lambda_1
      = \left(1+\frac{p}{q}\right) \left(G + \mathcal{D} \right) - \frac{p}{q},\label{eq:La2}
\end{eqnarray}
where $\mathcal{D}$ is the (renormalized) angular momentum deficit \citep[AMD, e.g.][]{laskar_spacing_2000} defined by:
\begin{equation}
  \mathcal{D} = \sum_{i=1}^2 x_i \overline{x}_i\quad (= \Lambda_1 + \Lambda_2 - G).
\end{equation}

Therefore, the Keplerian part is obtained by substituting $\Lambda_i$ values in the expression of $\mathcal{H}_0$.
This expression is then expanded to a given degree in eccentricities (i.e. to a given power in $x_i$ variables).
Note that the Keplerian part is now expressed as a power series of eccentricities even if it only depends
on semi-major axes in Delaunay coordinates system.
It should be remarked that $x_i$ variables only appear in the Keplerian part with $\mathcal{D}$
(which is of degree two and symmetrical for both planets).
As a consequence the Keplerian part only contains terms of even degree in eccentricities.

For the perturbative part, we use the expansion method presented in \citet{laskar_stability_1995}
and implemented in the algebraic manipulator TRIP \citep{gastineau_trip_2011}.
The Poincar\'e coordinates used in \citet{laskar_stability_1995}, noted $x$, and $x'$, are related
to our rectangular coordinates with the relations:
\begin{eqnarray}
  x\ &=& \overline{\scaled{x}}_1\ \expo{\ImUnit \sigma_0},\\
  x' &=& \overline{\scaled{x}}_2\ \expo{\ImUnit \sigma_0},
\end{eqnarray}
with:
\begin{equation}
  \sigma_0 = -\frac{p}{q} \lambda_1 + \left(1+\frac{p}{q}\right)\lambda_2.
\end{equation}

\citet{laskar_stability_1995} method is aimed at computing the perturbative part of the Hamiltonian in power series of $x$, $x'$
and in Fourier series of the mean longitudes $\lambda_1$, $\lambda_2$ (with coefficients depending on $\scaled{\Lambda}_i$).
The averaging method described in section \ref{sec:averaging} consists in selecting in this expansion
the terms that do not depend upon combinations of the mean longitudes other than $\sigma_0$ (and its harmonics).

Then it is straightforward to express $\mathcal{H}_1$ as a function of $x_i$ and $G$, by substituting $x$, $x'$ and $\scaled{\Lambda}_i$ by their values and renormalizing by $\Gamma$.
We obtain a power series of $x_i$ with coefficients depending on $G$.

The ratio between the perturbative and the Keplerian parts is of the order of
the mass ratio between the planets and the star.
To be consistent, the Keplerian part must be expanded to a higher degree in eccentricities than
the perturbative part.

As an example, for a first order resonance (of the form $(p+1)$:$p$), the Hamiltonian expanded to the fourth degree in eccentricities for the Keplerian part and to the second degree for the perturbation (noted degree 4-2) has the form:
\begin{equation}\label{eq:resham}
  \begin{array}{l c l}
    \mathcal{H} &=& K^{(2)}\mathcal{D}
      + K^{(4)} \mathcal{D}^2\\
      &+& S^{(0)}\mathcal{D}
      + S_1^{(2)} x_1\overline{x}_1 + S_2^{(2)} x_2\overline{x}_2
      + S_{12}^{(2)} (x_1\overline{x}_2 + x_2\overline{x}_1)\\
      &+& R_1^{(1)} (x_1 + \overline{x}_1) + R_2^{(1)} (x_2 + \overline{x}_2)\\
      &+& R_1^{(2)} (x_1^2 + \overline{x}_1^2) + R_2^{(2)} (x_2^2 + \overline{x}_2^2)
      + R_{12}^{(2)} (x_1 x_2 + \overline{x}_1\overline{x}_2),
  \end{array}
\end{equation}
where all coefficients ($K$, $S$, $R$) depends on $G$ and on the masses, and $K$ stands for Keplerian terms,
$S$ for secular terms and $R$ for resonant terms (see appendix \ref{app:calcoef} for a description of their computation).
For a resonance of order two (or more),
the expression would be similar but there would not be resonant terms of the degree one.
These terms are very important in the dynamics since they introduce constant terms (non-zero right-hand side)
in Hamilton's equations (\refeq{eq:Hameq}).
When there are no first degree terms (resonance of order two or more),
$x_i = 0$ ($i=1,2$) is always a fixed point.
On the contrary, for first order resonances, this fixed point does not exist and an initially circular system will
not stay circular.
Note that since the Hamiltonian is expanded as series of eccentricities, this model is only valid for low eccentricities.

\subsection{Study of the dynamics}
\subsubsection{Energy levels}
We consider a first order resonance and construct the resonant normal form
developed up to degree 4 for the Keplerian part and to first degree for the perturbation (noted degree 4-1).
For given values of the masses of the three bodies, a given mean-motion commensurability $(p+q)$:$p$,
and a given value of $G$, we are able to compute the needed coefficients $K$, $S$ and $R$.
We can then look at the energy levels of this Hamiltonian.
Since our problem has two degrees of freedom and four dimensions,
we cannot have a global view of these energy levels but we can represent them on section planes.
Figure~\ref{Figa} shows the energy levels, represented in different section planes (see the legend),
in the case of the 2:1 MMR
with a star mass of $m_0 = m_\sun$ (solar mass), planets masses of $m_1 = m_2 = m_\earth$ (Earth's mass),
and with $G = G_{0} (1-10^{-4})$.
$G_0$ is the value of the total circular angular momentum at nominal resonance.
At nominal resonance, the exact commensurability of \modif{mean-motions} induces the relation:
\begin{eqnarray}
  \frac{n_{1,r}}{n_{2,r}} &=& 1+\frac{q}{p} = \left(\frac{\Lambda_{2,r}}{\Lambda_{1,r}}\right)^3 \eta^3, \\
  \eta &=& \left(\frac{\mu_1}{\mu_2}\right)^{2/3} \frac{\beta_1}{\beta_2}
  = \left(\frac{m_0+m_2}{m_0+m_1}\right)^{1/3} \frac{m_1}{m_2}.
\end{eqnarray}
Together with the relation $\left(1 + \frac{q}{p}\right) \Lambda_{1,r} + \Lambda_{2,r} = 1$ (see \refeq{eq:La2}), this imposes the values $\Lambda_{1,r}$
and $\Lambda_{2,r}$ at nominal resonance:
\begin{eqnarray}
  \Lambda_{1,r} & = & \left[
  \left(1+\frac{q}{p}\right)
  + \left(1+\frac{q}{p}\right)^{1/3} \eta^{-1} \right]^{-1},\\
  \Lambda_{2,r} & = & \left[
  1 + \left(1+\frac{q}{p}\right)^{2/3} \eta \right]^{-1},
\end{eqnarray}

$G_0$ is then given by:
\begin{equation}
  G_0 = \Lambda_{1,r} + \Lambda_{2,r}
  = \left\{ 1 + \frac{q}{p} \left[
  1 + \left( 1 + \frac{q}{p}\right)^{1/3} \eta^{-1} \right]^{-1}
  \right\}^{-1}.
\end{equation}

\figa

\figb

On Fig.~\ref{Figa}, we distinguish three areas in the different section planes
(except in the plane $(\sin\sigma_1,\sin\sigma_2)$ on the top right of the graph).
There are two zones of circulation:
internal circulation for low eccentricities and external circulation for high eccentricities.
Between these two circulation areas, we observe a libration zone (banana-shaped level curves)
separated from both circulation areas by two separatrices.
Fixed points of the system (see next section) are marked with colored dots (for stable ones)
and crosses (for unstable ones).
The red dot corresponds to the libration center in Fig.~\ref{Figa}, bottom-left.

Figure~\ref{Figb} shows the energy levels on the section plane defined by $\sin\sigma_1=\sin\sigma_2=0$
for increasing values of the parameter $G$ and for the 2:1 and the 3:2 MMR.
The masses are the same as for Fig.~\ref{Figa} ($m_0 = m_\sun$, $m_1 = m_2 = m_\earth$).
Figure~\ref{Figb},C1 is the same as Fig.~\ref{Figa}, bottom-left.
When $G$ decreases (Figs.~\ref{Figb},A1,B1), the libration area moves to higher eccentricities
and the internal circulation area takes more space.
On the contrary, for higher values of $G$ (Figs.~\ref{Figb},D1 to F1), the internal circulation and the libration areas
tend to shrink and eventually completely disappear leaving only the external circulation area.

We observe the same evolution for the 3:2 MMR (Figs.~\ref{Figb},A2 to F2) and it seems that this behavior is common
to all first order MMR.

Even if these section planes give an insight on the nature of the motion,
it is important to keep in mind that the real motion occurs on the total four dimensional phase space
and that these sections cannot provide a global picture of the dynamics.
The easiest way to have a good insight on the global structure of the phase space is
to look at the positions and natures of the fixed points of the system.
These fixed points are commonly referred to as apsidal corotation resonances or ACR \citep[e.g.][]{ferraz-mello_symmetrical_1993}.

\subsubsection{Fixed points}
The positions of the fixed points are obtained solving Hamilton's equations (\refeq{eq:Hameq}).
For instance, at degree 4-2 (\refeq{eq:resham}):
\begin{eqnarray}
  0 &=& \left( K^{(2)}
      + 2 K^{(4)} \mathcal{D}
      + S^{(0)}
      + S_1^{(2)}\right) x_1
      + S_{12}^{(2)} x_2 \nonumber\\
      && + 2 R_1^{(2)} \overline{x}_1
      + R_{12}^{(2)} \overline{x}_2
      + R_1^{(1)}, \label{eq:acr1}\\
  0 &=& S_{12}^{(2)} x_1
      + \left( K^{(2)}
      + 2 K^{(4)} \mathcal{D}
      + S^{(0)}
      + S_2^{(2)}\right) x_2 \nonumber\\
      && + R_{12}^{(2)} \overline{x}_1
      + 2 R_2^{(2)} \overline{x}_2
      + R_2^{(1)}. \label{eq:acr2}
\end{eqnarray}
This imply to solve a system of four polynomial real equations (real and imaginary parts of
Eqs.~(\ref{eq:acr1}), (\ref{eq:acr2})) with four real unknowns (real and imaginary parts of $x_1$, $x_2$).
We solve it using the \textit{Maple RootFinding[Isolate]} function \citep[see][]{rouillier_solving_1999}.
Then, for each solution, we linearize the equations of motion around the fixed point in order to
compute the eigenvalues and to distinguish between elliptic (purely imaginary eigenvalues)
and hyperbolic (nonzero real parts) fixed points.

\paragraph{Degree 4-1\\}
\figc
Figure~\ref{Figc} shows the positions and nature of ACR solutions as functions of $G$,
in case of the 2:1 MMR with the same masses than before.
The Hamiltonian is still developed up to degree 4-1.
On this figure, continuous lines represent stable fixed points and dashed lines represent
unstable ones. The choice of colors is consistent with Figs.~\ref{Figa} and \ref{Figb}.
We gave the positions of ACR only in the directions $e_1\cos\sigma_1$ and $e_2\cos\sigma_2$
because all ACR solutions that we found have $\sin\sigma_1=\sin\sigma_2=0$.
On the left of Fig.~\ref{Figc} (lowest values of $G$), there are two elliptic and one hyperbolic ACR.
This corresponds to the situation of Figs.~\ref{Figa} and \ref{Figb},A1 to D1.
The red elliptic ACR corresponds to the center of the libration area.
The green one corresponds to the center of the internal circulation area.
The blue hyperbolic ACR corresponds to the crossing of internal and external separatrices.

Around $G=G_0$, we observe a bifurcation and the green and the blue ACR disappear whereas
the red one remains the only one to survive and it stay stable.
This corresponds to the situation observed in Fig.~\ref{Figb},E1,F1 and which cannot be considered
as a resonant situation.

In order to have a better view of the spatial positions of these ACR we plotted on Fig.~\ref{Figd} the semi-major
axes ratio of ACR solutions as functions of $G$.
We observe that the center of libration corresponds to values of $\alpha$ lower than the nominal resonant value.
This is equivalent to values of $P_2/P_1$ greater than the nominal resonant value ($2$).
It means that planets are farther away from each other than the nominal resonance distance.
On the contrary, the hyperbolic ACR and the center of the internal resonant area correspond to higher values of $\alpha$,
thus two planets closer to each other than the nominal resonance distance.

Note that when the system moves away from the semi-major axes nominal resonant ratio,
both elliptic ACR solutions tend to zero eccentricities (Figs.~\ref{Figc} and \ref{Figd}).

\figd

\paragraph{Higher degrees of development\\}
\fige

In order to evaluate the upper bound of the eccentricities for which
the situation that we described is realistic we compare the results
obtained for increasing degree of development of the Hamitonian.
If the global structure of the phase space remains the same for higher degrees and the
number, natures and positions of ACR seem to converge we can consider that our model
is realistic.

We realized different tests in order to compare the structure of the phase space for different degrees.
It seems that the degree of development of the Keplerian part is not significantly affecting the structure
of the phase space as long as we keep the first four degrees terms. The main effect of the subsequent terms
is to slightly shift the structures.
However, the degree of development of the perturbative part is more important.

Figures~\ref{Fige}, and \ref{Figf} show the ACR positions of the 2:1 MMR as functions of $G$
in the same case than before but with a resonant Hamiltonian developed up to degree 4-2, 4-3, and 4-4.
Figure~\ref{Fige} show the ACR positions in the directions of $\cos\sigma_i$, and Fig.~\ref{Figf} show them in
the directions of $\sin\sigma_i$.
We only plot the degree 4-2 in Fig.\ref{Figf} because at degrees 4-3, and 4-4, all ACR solutions have
$\sin\sigma_i = 0$.

We see that around $G=G_0$ the structure of the phase space remains the same as for degree 4-1.
However, the presence of degree two terms induces a significant change in the position of the libration center
in the direction $e_2 \cos\sigma_2$. This shift is also observed at degrees 4-3 and 4-4.
For $G<G_0(1-2\times10^{-3})$ a more complex structure appears at degree 4-2 with two new fixed points that
diverge from the libration center in the direction of $\sin\sigma_i$.
Note that this structure disappears at degrees 4-3 and 4-4.
We thus interpret this structure as an artifact due to a too low degree of development.
Between degree 4-3 and degree 4-4 there are no significant changes in the structure of the phase space
but the degree 4-4 brings small corrections to the positions of ACR solutions.
We thus conclude that for the small eccentricities considered here, the structure would no change
qualitatively if we consider higher degrees of development.

\figf

Moreover, the structure of the phase space at higher eccentricities is not important for the following discussion
since we are interested in the very end of a tidal circularization process of a resonant system
when planets have low eccentricities.

\figg

\figh

The same analysis can be done for any first order MMR.
For instance, Figs.~\ref{Figg}, and \ref{Figh}
show the ACR positions of the 3:2 MMR in the same conditions (masses, etc.) as previously
and with a Hamiltonian developed up to degree 4-1 (Fig.\ref{Figg}), 4-2, 4-3, and 4-4 (Fig.\ref{Figh}).
The structure of the phase space is very similar at low eccentricities.
As for the 2:1 resonance, at degree 4-2, some complex structures appear
for low values of the parameter $G$.
However these structures are completely different between degrees 4-2, 4-3, and 4-4 so they should
also be consider as artifacts.
For $G>G_0(1-1.5\times10^{-3})$ the structure of the phase space is the same between degrees 4-3 and 4-4,
and is very similar to the structure observed for the 2:1 MMR.
This should not change much if we consider higher degrees of development and this structure seems to be
common to all first order MMR.

\section{Dynamics of two resonant planets with dissipation}\label{sec:dissipation}
In this section we consider the case of two planets in resonance in the presence of a dissipative force
(tidal effect with the star, planet-disk interactions, etc.).
In this case, $\scaled{G}$ and $\Gamma$ are no longer constants of the motion.
However, if the dissipative force is sufficiently weak, their evolution is slow and we can apply
the adiabatic invariant theory \citep[see][]{henrard_adiabatic_1982, henrard_second_1983}.
This means that the short term evolution of the system is still close to the one observed in the conservative case.
On the long term, the dissipation affects the constants $\scaled{G}$ and $\Gamma$ and thus the parameter
$G$.
When $G$ evolves, the phase space of the system evolves as presented in the previous section.

Depending on the type of dissipation, $G$ can either increase or decrease.
Its evolution depends on how the dissipation affects the eccentricities and the semi-major axes of both planets.
More precisely, the derivative of $G$ is given by:
\begin{equation}\label{eq:dotG}
  \begin{array}{l l}\displaystyle
    \left.\deriv{G}{t}\right|_\mathrm{d} =
      &- \Lambda_1 \Frac{e_1^2}{\sqrt{1-e_1^2}} \left.\left(\Frac{\dot{e}_1}{e_1}\right)\right|_\mathrm{d}\\
      &- \Lambda_2 \Frac{e_2^2}{\sqrt{1-e_2^2}} \left.\left(\Frac{\dot{e}_2}{e_2}\right)\right|_\mathrm{d}\\
      &+ \Frac{1}{2} \Lambda_1 \Lambda_2 \left( \sqrt{1-e_1^2} - \left(1+\Frac{q}{p}\right)
        \sqrt{1-e_2^2}\right) \left.\left(\Frac{\dot{\alpha}}{\alpha}\right)\right|_\mathrm{d},
  \end{array}
\end{equation}
where $\left.\left(\dot{e}_i/e_i\right)\right|_\mathrm{d}$, $\left.\left(\dot{\alpha}/\alpha\right)\right|_\mathrm{d}$ are given by the dissipation model and $\Lambda_i$ are the renormalized (by $\Gamma$) actions.

\subsection{Migration}
In the case of a strong migration of the planets and with low eccentricities,
the dominant term in \refeq{eq:dotG} is the third one which depend on the semi-major axes ratio evolution.
For low eccentricities we have:
\begin{equation}
  \sqrt{1-e_1^2} - \left(1+\frac{q}{p}\right)\sqrt{1-e_2^2} \sim -\frac{q}{p} < 0.
\end{equation}
Thus, if the planets undergo convergent migration
$\left(\left.\left(\dot{\alpha}/\alpha\right)\right|_\mathrm{d} > 0\right)$,
$G$ should decrease
and if the migration is divergent, $G$ should increase.
As it is already known \citep[see][]{henrard_second_1983}, a resonant capture can occur only if the migration is convergent
and the resonance is crossed with a decreasing parameter $G$.

\subsection{Tidal circularization}
\subsubsection{Evolution of the parameter $G$}
In the case of circularization of the orbits due to tides raised on the planets by the star,
the angular momentum of each planet is almost conserved and the dissipation in eccentricities reads to lower order in
eccentricities \citep[see][]{papaloizou_tidal_2011}:
\begin{equation}\label{eq:dote}
  \left.\deriv{e_i}{t}\right|_\mathrm{d} = - \frac{e_i}{t_{c, i}},
\end{equation}
where $t_{c,i}$ is a damping constant (see \refeq{eq:tcirc}).
Since the angular momentum of each planet remains unaffected by the dissipation,
the semi-major axes compensate the eccentricities decreases.
The evolution of semi-major axes then reads to lowest order in eccentricities:
\begin{equation}\label{eq:dota}
  \left.\deriv{a_i}{t}\right|_\mathrm{d} =  2 a_i e_i \left.\deriv{e_i}{t}\right|_\mathrm{d}
  = - 2 e_i^2 \frac{a_i}{t_{c, i}}.
\end{equation}

In this case, the easiest way to compute how the dissipation affects the constant $G$ is not to use
\refeq{eq:dotG} but to go back to the definition of $G$.
Since $G=\scaled{G}/\Gamma$ and $\scaled{G}$ is conserved, we have:
\begin{equation}
  \left.\left(\frac{\dot{G}}{G}\right)\right|_\mathrm{d} =
  - \left.\left(\frac{\dot{\Gamma}}{\Gamma}\right)\right|_\mathrm{d}.
\end{equation}
From \refeq{eq:dota} we deduce the impact of the circularization process on $\scaled{\Lambda}_i$
(to lowest order in eccentricities):
\begin{equation}
  \left.\deriv{\scaled{\Lambda}_i}{t}\right|_\mathrm{d} = - e_i^2 \frac{\scaled{\Lambda}_i}{t_{c, i}}.
\end{equation}
Therefore, the evolution of $\Gamma$ is given by:
\begin{equation}\label{eq:dotGamma}
  \left.\deriv{\Gamma}{t}\right|_\mathrm{d} = -  \left(1 + \frac{q}{p}\right) e_1^2 \frac{\scaled{\Lambda}_1}{t_{c, 1}}
   - e_2^2 \frac{\scaled{\Lambda}_2}{t_{c, 2}} < 0.
\end{equation}
And the evolution of $G$ is governed by:
\begin{equation}\label{eq:dotG2}
  \left.\deriv{G}{t}\right|_\mathrm{d} = G \left( \left(1 + \frac{q}{p}\right) e_1^2 \frac{\Lambda_1}{t_{c, 1}}
   + e_2^2 \frac{\Lambda_2}{t_{c, 2}} \right) > 0.
\end{equation}

It results that, in the case of tidal circularization of the orbits, $G$ slowly increases with time
(dominant term of order 2 in eccentricities).
Looking at Figs.~\ref{Figc} to \ref{Figh}, the position of the system on these graphs
will slowly migrate from the left to the right, and eventually,
the system will pass through the bifurcation (around $G=G_0$).
The only remaining ACR solution is the one that corresponds to the libration center
but
\modif{the separatrix of the resonance and the resonant area disappear.}
The motion around the only remaining fixed point is elliptic and
this fixed point is quickly decreasing to zero eccentricities.
Thus, the motion should be very well approximated by the secular problem
(using a complete averaging instead of the partial one introduced to obtain the resonant normal form).
Looking at Fig.~\ref{Figd}, we see that the remaining ACR depart from exact commensurability
with a lower value of $\alpha$ (and higher value of $P_2/P_1$) than the nominal resonance value.
This corresponds to planets farther away from each other than the nominal resonant distance.
Therefore, as noticed by \citet{papaloizou_dynamics_2010}
the long-term effect of the tidal circularization is a repulsion of both planets.
This is why \citet{batygin_dissipative_2012} and \citet{lithwick_resonant_2012} invoke this process to
explain the excess of Kepler systems whose planets are just slightly further away from each other than nominal
resonances.

\subsubsection{Motion around the libration center}
The dissipation strongly affects the eccentricities (dominant terms of order one in \refeq{eq:dote})
whereas the constants $\Gamma$ (or $G$) are only slightly affected (dominant terms of order two in
Eqs.~(\ref{eq:dotGamma}),(\ref{eq:dotG2})).
Thus, we can consider that the damping of eccentricities happens on a shorter time scale than the evolution
of $\Gamma$ and $G$.
These quantities can thus still be considered as constants (in first approximation)
when we take into account the damping of eccentricities
in the equations of motion (adiabatic invariant theory).
From \refeq{eq:dote}, we deduce:
\begin{equation}\label{eq:xdissip}
  \left.\deriv{x_i}{\tau}\right|_\mathrm{d} = - \frac{x_i}{\tau_{c, i}},
\end{equation}
with $\displaystyle \tau_{c, i} = \frac{t_{c, i}}{\Gamma^3}$.

Now, suppose that the system lie in the vicinity of the libration center of a first order MMR
whose position is noted $x^0_i$ ($i=1,2$).
Let us define the vector $\vec{x}$ as:
\begin{equation}
  \vec{x} =
  \left( \begin{array}{ c }
           x_1\\
           x_2\\
           \overline{x}_1\\
           \overline{x}_2
         \end{array} \right).
\end{equation}

We can linearize the equations of motion (of the conservative case) around the fixed elliptic point $\vec{x}^0$:
\begin{equation}
  \dot{\vec{x}} = \deriv{\vec{x}}{\tau} \approx \ImUnit A_{\vec{x}^0} \left(\vec{x} - \vec{x}^0\right).
\end{equation}

Equations of motion of the dissipative problem are thus simply given (to first order in eccentricities) by:
\begin{equation}
  \dot{\vec{x}} \approx \ImUnit A_{\vec{x}^0} \left(\vec{x} - \vec{x}^0\right) - \delta B \vec{x},
\end{equation}
with (\refeq{eq:xdissip}):
\begin{equation}
  \delta B = 
  \left( \begin{array}{ c c c c }
    \frac{1}{\tau_{c, 1}} & 0 & 0 & 0\\
    0 & \frac{1}{\tau_{c, 2}} & 0 & 0\\
    0 & 0 & \frac{1}{\tau_{c, 1}} & 0\\
    0 & 0 & 0 & \frac{1}{\tau_{c, 2}}
  \end{array} \right).
\end{equation}
Thus:
\begin{equation}
  \dot{\vec{x}} = \left(\ImUnit A - \delta B\right) \left( \vec{x} - \vec{x}^1 \right),\label{eq:newlineq}
\end{equation}
with
\begin{equation}
  \vec{x}^1 = \left(\ImUnit A - \delta B\right)^{-1} \ImUnit A \vec{x}^0 = \left(\Id + \ImUnit A^{-1} \delta B\right)^{-1}.
\end{equation}
Since $\delta B$ is a pertubation of $\ImUnit A$,
we can make the approximation (to first order):
\begin{equation}
  \left( \Id + \ImUnit A^{-1} \delta B \right)^{-1}
  \approx \left( \Id - \ImUnit A^{-1} \delta B\right) .
\end{equation}
Hence:
\begin{equation}\label{eq:newpos}
  \vec{x}^1 \approx \left( \Id - \ImUnit A^{-1} \delta B\right) \vec{x}^0 = \vec{x}^0 - \ImUnit A^{-1} \delta B \vec{x}^0.
\end{equation}

The dynamics around the libration center is thus modified by the dissipation in two different ways.
First, the position of the fixed point is slightly changed by the offset $A^{-1} \delta B \vec{x}^0$
in the imaginary directions (see \refeq{eq:newpos}).
Secondly, the matrix giving the motion around the fixed point is modified (see \refeq{eq:newlineq}).

Noting $\vec{y} = \vec{x} - \vec{x}^1$, we can apply directly the method described in \citet{laskar_tidal_2012} to $\vec{y}$.
This method highlights the effects of the pertubation on the eigenvectors (i.e. the diagonalizing linear transformation) and
on the eigenvalues of the system.
If we note $S_0$ the diagonalizing matrix and $D_0 = \textrm{diag}(g_1,g_2,-g_1,-g_2)$ the diagonalized matrix of the conservative case such as:
\begin{eqnarray}
  \vec{y} &=& S_0 \vec{u},\\
  D_0 &=& S_0^{-1} A S_0,\\
  \dot{\vec{u}} &=& \ImUnit D_0 \vec{u}\label{eq:propmod},
\end{eqnarray}
where $\vec{u}$ is the vector of eigenmodes and $g_1$, $g_2$ are the eigenvalues.
The diagonalizing matrix and the diagonalized matrix in the dissipative case are given by:
\begin{eqnarray}
  S &=& S_0 ( \Id + \ImUnit \delta S_1),\\
  D &=& \ImUnit D_0 - \delta D_1,
\end{eqnarray}
The small change $\delta S_1$ of the eigenvectors \citep[see][]{laskar_tidal_2012} does not introduce a major change in
the dynamics but can be seen as a corrective term as well as the correction on the position of the fixed point.
However, the pertubation $\delta D_1$ of the eigenvalues is much more important.
Noting $\delta D_1 = \textrm{diag}(\gamma_1,\gamma_2,\gamma_1,\gamma_2)$, we have:
\begin{equation}
  \gamma_i = \left(S_0^{-1} \delta B S_0\right)_{ii},
\end{equation}
and these coefficients are real and positive.
Finally the diagonalized equations of motion now read:
\begin{equation}
  \dot{\vec{u}} = \textrm{diag}(\ImUnit g_1 - \gamma_1, \ImUnit g_2 - \gamma_2, -\ImUnit g_1 - \gamma_1, -\ImUnit g_2 - \gamma_2)\vec{u}.
\end{equation}

Thus, the dissipation introduces negative real parts (-$\gamma_i$) in the eigenvalues of the system.
Therefore, all the eigenmodes will be damped on time scales given by these coefficients
and the fixed point is no more elliptic but attractive.
The main short-term effect of the dissipation is to induce an attraction of the system towards the libration center.

To sum up, the tidal circularization process induce two main effects for planets that are initially resonant.
On the short-term, planets tend to reach the stable ACR solution corresponding to the libration center
and whose position is slightly modified by the dissipative terms.
On the long-term, planets tend to leave the resonance by moving away from each other
(the parameter $G$ increases and the planets follow the stable ACR outside of the resonance).

Note that, even outside of the resonance, the attractive ACR solution continues to exist and
is not exactly at zero eccentricities.
We will focus in the next section on the behavior of the resonant angles outside the resonances.

\section{Secular motion at very low eccentricities}\label{sec:secular}
In this section we consider a conservative system of two planets around a star (on the same plane)
that are far from any resonance and which have very low eccentricities.
More precisely, we suppose that eccentricities have already been damped by the dissipation and we study the dynamics
of the system after the dissipation process.
We are in the field of application of the secular normal form.
We start our study with the Poincar\'e rectangular astrocentric coordinates:
$(\lambda_i, \Lambda_i, y_i, -\ImUnit \overline{y}_i)$ for each planet ($i=1,2$) \citep[e.g.][]{laskar_stability_1995}.
Note that, for the simplicity of notations, we omit the hats even if these variables are not renormalized anymore by $\Gamma$.
$y_1$, $y_2$ are the usual Poincar\'e rectangular coordinates that are noted $x$, $x'$ in \citet{laskar_stability_1995}:
\begin{equation}\label{eq:poinca}
  y_i = \sqrt{\Lambda_i}\sqrt{1-\sqrt{1-e_i^2}} \expo{\ImUnit \omega_i} = \bar{\scaled{x}}_i \expo{\ImUnit \sigma_0}.
\end{equation}

The Hamiltonian can be developed in power series of $y_i$, $\overline{y}_i$ and in Fourier series
of the mean longitudes $\lambda_i$:
\begin{eqnarray}
  \mathcal{H} &=& \mathcal{H}_0(\Lambda)
    + \mathcal{H}_1 (\lambda,\Lambda,y,\overline{y}),\\
  \mathcal{H}_1 &=& \sum_{k,m,\overline{m}} h_{k,m,\overline{m}} (\Lambda)
  y_1^{m_1} \overline{y}_1^{\overline{m}_1} y_2^{m_2} \overline{y}_2^{\overline{m}_2}
  \expo{\ImUnit (k_1 \lambda_1 + k_2 \lambda_2)},
\end{eqnarray}
where the D'Alembert rule reads:
\begin{equation}
  k_1 + k_2 + m_1 + m_2 - \overline{m}_1 - \overline{m}_2 = 0,
\end{equation}
and the coefficients $h_{k,m,\overline{m}}$ are functions of $\Lambda_1$, $\Lambda_2$ which can be expressed in
terms of Laplace coefficients \citep[see][]{laskar_stability_1995}.

To first order of the planets masses, the secular normal form is obtained by averaging the perturbative part
of the Hamiltonian over the mean longitudes.
This is performed by introducing a change of coordinates close to the identity.
Let us note $(\secul{\lambda}_i, \secul{\Lambda}_i, \secul{y}_i, -\ImUnit \overline{\secul{y}}_i)$ the new coordinates,
$\secul{\mathcal{H}}$ the new Hamiltonian, and $W_1$ the generating Hamiltonian of the transformation.
By definition, if we note $\secul{\mathcal{H}}_0$ and $\secul{\mathcal{H}}_1$ the Keplerian part and the first order
part (in planets masses) of the new Hamiltonian, the transformation reads (to first order in planets masses):
\begin{eqnarray}
  \secul{\mathcal{H}}_0 &=& \mathcal{H}_0,\\
  \secul{\mathcal{H}}_1 &=& \mathcal{H}_1 + \{W_1, \mathcal{H}_0\},
\end{eqnarray}
where the Poisson brackets are noted with braces.
The secular normal form (at the first order) is obtained by imposing:
\begin{equation}
  \secul{\mathcal{H}}_1 = \mathcal{H}_1 + \{W_1, \mathcal{H}_0\} =  \langle \mathcal{H}_1 \rangle.\label{eq:homol}
\end{equation}
\refeq{eq:homol} is commonly called the homological equation and its solution is given by:
\begin{equation}
  W_1 = \sum_{k\neq(0,0),m,\overline{m}} \frac{ h_{k,m,\overline{m}} (\secul{\Lambda}) }{\ImUnit(n_1 k_1 + n_2 k_2)}
  \secul{y}_1^{m_1} \overline{\secul{y}}_1^{\overline{m}_1} \secul{y}_2^{m_2} \overline{\secul{y}}_2^{\overline{m}_2}
  \expo{\ImUnit (k_1 \secul{\lambda}_1 + k_2 \secul{\lambda}_2)},
\end{equation}
where $n_1$, $n_2$ are the unperturbed Keplerian mean-motions of the planets.

By construction the secular Hamiltonian reads to first order of the masses:
\begin{equation}
  \secul{\mathcal{H}} = \mathcal{H}_0 (\secul{\Lambda}) + \sum_{m,\overline{m}} h_{(0,0),m,\overline{m}} (\secul{\Lambda})
  \secul{y}_1^{m_1} \overline{\secul{y}}_1^{\overline{m}_1} \secul{y}_2^{m_2} \overline{\secul{y}}_2^{\overline{m}_2}.
\end{equation}
$\secul{\Lambda}_1$ and $\secul{\Lambda}_2$ are constants of the motion (first integrals) and $\secul{y}_1 = \secul{y}_2 = 0$ is a stable fixed point around which the equations of motion can be linearized (Lagrange-Laplace theory).
The change of variables is constructed to be close to the identity, but the reversion to the initial
set of coordinates reintroduces short period terms.
This reversion reads up to the first order in planets masses:
\begin{eqnarray}
  \lambda_i &=& \secul{\lambda}_i + \{ W_1, \secul{\lambda}_i \} = \secul{\lambda}_i + \deriv{W_1}{\secul{\Lambda}_i},\\
  \Lambda_i &=& \secul{\Lambda}_i + \{ W_1, \secul{\Lambda}_i \} = \secul{\Lambda}_i - \deriv{W_1}{\secul{\lambda}_i},\\
  y_i &=& \secul{y}_i + \{ W_1, \secul{y}_i \} = \secul{y}_i + \ImUnit \deriv{W_1}{\overline{\secul{y}}_i},
\end{eqnarray}
In general, this reversion to the original set of variables only introduces small corrections
to the dominant secular terms.
But if secular eigenmodes are totally damped (i.e. $\secul{y}_1 = \secul{y}_2 = 0$),
the only terms that are nonzero in the expressions of $y_1$, $y_2$ are those corresponding
to first order mean-motion commensurabilities:
\begin{eqnarray}
  y_1 &=& \ImUnit \deriv{W_1}{\overline{\secul{y}}_1}
  = \sum_{p} \frac{ h_{(-p,p+1),(0,0),(1,0)} (\secul{\Lambda}) }{n_2 (p+1) - n_1 p}
   \expo{\ImUnit \left((p+1) \secul{\lambda}_2 - p \secul{\lambda}_1\right)},\\
  y_2 &=& \ImUnit \deriv{W_1}{\overline{\secul{y}}_2}
  = \sum_{p} \frac{ h_{(-p,p+1),(0,0),(0,1)} (\secul{\Lambda}) }{n_2 (p+1) - n_1 p}
   \expo{\ImUnit \left((p+1) \secul{\lambda}_2 - p \secul{\lambda}_1\right)}.
\end{eqnarray}

\fign

Now, if we consider a system which is near a first order mean-motion commensurability $(p+1)$:$p$ but still outside it,
the divisor $(p+1) n_2 - p n_1$ is smaller than the others (but not considerably small)
and the corresponding term dominates the evolution of $y_1$ and $y_2$.
Thus $y_1$ and $y_2$ are of the form:
\begin{eqnarray}
  y_1 &\approx& \frac{ h_{(-p,p+1),(0,0),(1,0)} (\secul{\Lambda}) }{n_2 (p+1) - n_1 p}
   \expo{\ImUnit \left((p+1) \secul{\lambda}_2 - p \secul{\lambda}_1\right)}, \label{eq:y1}\\
  y_2 &\approx& \frac{ h_{(-p,p+1),(0,0),(0,1)} (\secul{\Lambda}) }{n_2 (p+1) - n_1 p}
   \expo{\ImUnit \left((p+1) \secul{\lambda}_2 - p \secul{\lambda}_1\right)}. \label{eq:y2}
\end{eqnarray}
In terms of arguments of periastron (see \refeq{eq:poinca}), this means that:
\begin{equation}
  \omega_i \approx (p+1) \secul{\lambda}_2 - p \secul{\lambda}_1 + \epsilon_i,
\end{equation}
where $\epsilon_i = 0$ or $\pi$, depending on the sign of the amplitudes in Eqs.~(\ref{eq:y1}), (\ref{eq:y2}).
Of course, in this situation the resonant angles $\sigma_1 = (p+1) \lambda_2 - p \lambda_1 - \omega_1$ and
$\sigma_2 = (p+1) \lambda_2 - p \lambda_1 - \omega_2$ will librate (around $0$ or $\pi$).
Figure~\ref{Fign} illustrates this phenomenon of artificial libration.
It is important to notice that contrary to the resonant case, $(p+1) \lambda_2 - p \lambda_1$ and
$\omega_i$ are dominated by short periods (high frequencies).
In the resonant case, we always have $(p+1) n_2 - p n_1 \approx 0$,
thus $(p+1) \lambda_2 - p \lambda_1$ has a long period and $\omega_i$ are dominated by secular eigenmodes, which
also have long periods.
This is the reason why the classical secular averaging is valid in this case.

Thus, in the case of very low eccentricities, when eigenmodes are almost completely damped by dissipation (Fig.\ref{Fign}, A2,B2),
the fact that the resonant angles are librating (Fig.\ref{Fign}, B2) does not mean that the system is resonant.
It is just a geometrical effect due to the fact that the circulation center is not at zero. More precisely, the 
amplitude of the considered argument (Eqs.~(\ref{eq:y1}),(\ref{eq:y2}) ) is larger than the amplitude of the proper mode.

\section{Numerical simulation}\label{sec:simulation}
In order to confirm the different behaviors of resonant systems with dissipation that we highlighted in
section~\ref{sec:dissipation}, we ran two numerical simulations, one in the case of the 3:2 MMR and the other
for the 2:1 MMR.

\subsection{Simulation $S1$, 3:2 MMR}
The first numerical simulation concerns the two innermost planets of the
\object{GJ581} system which are near the 3:2 MMR \citep[e.g.][]{papaloizou_tidal_2011}.
Our simulation is comparable to the one presented in \citet{papaloizou_tidal_2011}.
The star mass is $m_0 = 0.31 m_\sun$, and planets masses are set to $m_1 = 1.94 m_\earth$ and
$m_2 = 15.64 m_\earth$. The initial semi-major axes are set to $a_1 = 0.11\ \mathrm{AU}$ and
$a_2 = 0.15\ \mathrm{AU}$.
Initial eccentricities are set to $e_1 = 0.01$, and $e_2 = 0.001$.
At the beginning of the simulation both planets undergo a migration process due to interactions with a disk.
We adopted the same migration time scale as \citet{papaloizou_tidal_2011}:
\begin{equation}
  t_{mig,i} = 4.375\times 10^5 \frac{m_\earth}{m_i}\ \mathrm{yr}.
\end{equation}
Because the outer planet is more massive, its migration is more efficient and the system undergo a convergent migration.
During this migration phase the time scale of the circularization is set to be $100$ times shorter than the migration
time scale \citep[see][]{lee_dynamics_2002}:
\begin{equation}
  t_{c,i} = \frac{t_{mig,i}}{100}.
\end{equation}
At $t = 10\ \mathrm{kyr}$, the migration stops (the disk is gone) and the only dissipative force
that remains is the tidal interaction with the star. The timescale of this circularization process is given by
(still following \citet{papaloizou_tidal_2011}):
\begin{equation}\label{eq:tcirc}
  t_{c,i} = 4.65\times 10^4 \left(\frac{m_\sun}{m_0}\right)^{(3/2)} \left(\frac{m_\earth}{m_i}\right)^{(2/3)}
  \left(\frac{20 a_i}{1\ \mathrm{AU}} \right)^{6.5} Q'\ \mathrm{yr},
\end{equation}
where $Q'$ is a parameter of the tidal dissipation model.
$Q'$ is set to $1.5$ in this simulation in order to reduce the computation time required
to see the effect of the circularization \citep[see][]{papaloizou_tidal_2011}.
A realistic value would be several order of magnitude greater.
This means that the time scale of the evolution of the system during the circularization process in our simulation
is much shorter than the realistic one.
We perform a direct n-body integration of the system using the ODEX integrator \citep[see][]{hairer_solving_2010}, with the dissipative force acting on each planet given by:
\begin{equation}
  \vec{F}_\mathrm{d} =
    - \frac{1}{t_{mig,i}} \deriv{\vec{r}_i}{t}
    - \frac{2}{r_i^2 t_{c,i}} \left(\deriv{\vec{r}_i}{t}.\vec{r}_i\right)\vec{r}_i,
\end{equation}

\figk

\figl

Figure~\ref{Figk} shows the evolution of semi-major axes, and eccentricities of both planets as well as the
evolution of the parameter $G$ at the beginning of the simulation (up to $20\ \mathrm{kyr}$).
During the first $10\ \mathrm{kyr}$ of the simulation, the planets undergo convergent migration and $G$ decreases
(Fig.~\ref{Figk},~bottom in gray).
The system enters very quickly in the 3:2 resonance and we can see on Fig.~\ref{Figk},~middle, that the semi-major
axes stay locked in the resonant ratio.
As explained by \citet{lee_dynamics_2002}, the eccentricities are excited by the resonance until they reach an equilibrium
state that depends on the strength of the circularization process which comes with the migration (Fig.~\ref{Figk},~top).
 
At $t = 10\ \mathrm{kyr}$, the migration stops and the only remaining dissipative force is
the tidal circularization of the orbits.
The long term evolution of $G$ is shown in Fig.~\ref{Figl}.
As expected, we see on Fig.~\ref{Figl} that the effect of the tidal circularization is an increase of $G$ that eventually
exceed $G_0$.
The increase is slower and slower, and at the end of the simulation $G$ is almost constant.

\figi

\figj

We can follow the evolution of the system on graphs similar to those of section~\ref{sec:conservative}.
Figure~\ref{Figi} gives the positions of 3:2 ACR solutions for the system considered in the simulation (masses),
superimposed over the successive positions of the system on these graphs.
The colors of the fixed points are the same as in section~\ref{sec:conservative}.
In particular, the center of libration of the resonant area is plotted in red.
The gray dots corresponds to positions of the system during the first $10\ \mathrm{kyr}$ (migration phase), whereas
the black dots corresponds to the circularization phase.
Since $G$ decreases during the migration phase (in gray) the system goes from the right to the left of the graphs.
On the contrary, during the tidal circularization phase (in black) $G$ increases and the system goes from the left
to the right of the graphs.

At the very beginning of the simulation, the system is not yet in the resonance.
On Fig.~\ref{Figi}, the system lie at the right of the bifurcation and there is only one fixed point.
The phase space is similar to the one of Figs.~\ref{Figb},~E2,F2.
Then, due to the convergent migration (gray dots) the system passes through the bifurcation and the phase space looks like
the one of Figs.~\ref{Figb},~A2 to D2. The system undergo oscillations around the libration center (red dot).
Finally, due to the circularization process (black dots), the system passes a second time through the bifurcation.
Thus, the system leaves the resonance and the phase space is again similar to Figs.~\ref{Figb},~E2,F2.

We see in Fig.~\ref{Figi} that the system follows very well the ACR solution corresponding to the libration center
both during the migration and the circularization phases.
We also see that the tidal dissipation induces a decrease of the amplitude of oscillations around this fixed point.
This corresponds to the expected damping of eigenmodes around the libration center (section~\ref{sec:dissipation}).

Figure~\ref{Figj} shows the evolution of the semi-major axes ratio (and periods ratio)
as a function of $G$ for the simulation
and for the fixed points of our model.
We see that during the migration process (in gray),
the semi-major axes ratio ($\alpha$) increases until it reaches the resonant ratio.
During this migration phase, $\alpha$ does not oscillate much, but when the system enter the resonance, $\alpha$ begins
to oscillate around the libration center position.
During the circularization process (in black), the system follows the libration center position, the oscillations are damped and the
semi-major axes ratio decreases (the periods ratio increases).
Thus the system quits the resonance with a periods ratio larger than the resonant one, as expected.

Note that when $G$ increases, the eccentricities quickly tend to zero (Fig.~\ref{Figi})
and the circularization process is less and less effective.
This is why $G$ increases slower and slower (Fig.~\ref{Figl}).
This means that the system cannot depart very far away from the MMR on a finite time
(see appendix~\ref{app:asympt} for an estimation of the asymptotic evolution of $G$).
This is even more true for a real system, since in this simulation the circularization process is several order of magnitude
more efficient than in reality.
This mechanism explains why Kepler systems lie just slightly farther away from the MMR.

Figure~\ref{Figm} shows the evolution of the resonant angle $\sigma_1$ (in red) and the difference of periastrons
$\Delta \omega = \sigma_2 - \sigma_1$ (in green).
We see that both resonant angles and $\Delta \omega$ librate during the whole simulation even if the system is
outside the resonance at the end of the simulation (see Figs.~\ref{Figi} and \ref{Figj}).
$\sigma_1$ librates around $0$, while $\sigma_2$ and $\Delta \omega$ librate around $\pi$.
This could have also been deduced from Fig.~\ref{Figi}.

\figm

\subsection{Simulation $S2$, 2:1 MMR}
\figkk

\figll

\figii

\figjj

\figmm

We also ran a simulation of the same system but in the case of the 2:1 MMR.
The masses of the star and the planets are the same but initial semi-major axes are set to
$a_1 = 0.11\ \mathrm{AU}$, and $a_2 = 0.18\ \mathrm{AU}$.
Initial eccentricities are set to $e_1 = 0.002$, and $e_2 = 0$.
All other parameters are the same as for the first simulation.

Figure~\ref{Figkk} shows the evolution of eccentricities, semi-major axes and $G$ during the first $20\ \mathrm{kyr}$
of the simulation.
Their behaviors are very similar to those of the simulation $S1$ (Fig.~\ref{Figk}).
Note that as for the resonance 3:2, $G$ decreases during the migration phase (in gray).
We plot in Fig.\ref{Figll} the long term evolution of $G$ which increases due to the circularization process.
Figures~\ref{Figii} and \ref{Figjj} show the superposition of the fixed points of our model with the successive positions
of the system.
Figure~\ref{Figii} gives the eccentricities as functions of $G$ and Fig.~\ref{Figjj} gives the semi-major axes ratio
(as well as the periods ratio) as a function of $G$.
As for the 3:2 resonance, in the case of the 2:1 resonance, the system begins outside the resonance
(on the right of the graphs) and the phase space corresponds to the situation of Figs.~\ref{Figb},~E1,F1.
Due to the convergent migration of planets the system goes from the right to the left (gray dots)
and passes through the bifurcation following the red ACR solution.
At this point the phase space is similar to those of Figs.\ref{Figb},~A1 to D1.
During the circularization phase, the system goes from the left to the right in Figs.~\ref{Figii}, \ref{Figjj} (black dots).
The system passes again \modif{at} the bifurcation and leaves the resonance (phase space of Figs.~\ref{Figb},~E1,F1).
Shortly after the system quited the resonance, the amplitude of oscillations around the fixed point suddendly increases in
the direction of $e_2$.
This is due to the crossing of a 2:1 resonance between the two proper modes of the motion around the
fixed point (see appendix~\ref{app:ressec} for more details).
However the dissipation damps again the oscillations after this event and the system keeps following the ACR solution.

We observe on Fig.~\ref{Figmm} that the first resonant angle $\sigma_1$ keeps librating around $0$
during the whole simulation whereas
$\Delta \omega$ and $\sigma_2$ circulate almost all the time.
More precisely, during the first $0.5\ \mathrm{Myr}$ of the simulation,
$\Delta \omega$ and $\sigma_2$ librate around $0$.
Then, these angles circulate during a short laps of time around $t=0.5\ \mathrm{Myr}$.
After this event, they begin to librate around $\pi$.
Actually, the system is still in resonance and oscillates around the libration center but
the position of this fixed point crosses the value $e_2 = 0$ for $G/G_0-1=5\times10^{-4}$ (see Fig.~\ref{Figii})
around $t=0.5\ \mathrm{Myr}$.
The dynamics does not change and this event is only geometrical.
Thus, even if the system is clearly in the resonance area, very close the the libration center, we observe a circulation of
$\sigma_2$.
Around $t=1.4\ \mathrm{Myr}$, $\Delta \omega$ and $\sigma_2$ begin to circulate (Fig.~\ref{Figmm}).
This corresponds to the crossing of the resonance (see appendix~\ref{app:ressec}) that we observe in Fig.~\ref{Figii}.
This event increases the amplitude of oscillations around the fixed point and due to this
higher amplitude the system passes through $e_2=0$ at each oscillation.
This is why we observe a circulation of $\Delta \omega$ and $\sigma_2$.

Note that both simulations show that it is possible to observe an oscillation of the resonant angles outside of resonances
($\sigma_1$ and $\sigma_2$ for $S1$, and only $\sigma_1$ for $S2$).
On the opposite, in simulation $S2$ we observe a circulation of $\sigma_2$ when
the libration center position crosses the axis $e_2=0$ while the system is
clearly inside the resonance and oscillates very close to the libration center.
Therefore, it appears that resonant angles have to be considered with great caution and cannot always be used
to distinguish between truly dynamical effects and simple geometrical effects.

\section{Conclusion}
In this work, we presented a study of planar resonant planetary systems in the conservative case and in presence
of a dissipative force.
Our main interest was to understand the dynamics at the end of a circularization process in resonant systems.
We used a completely analytical model developed in power series of eccentricities which is
well suited for the study of these low eccentricity systems.
Before introducing the dissipative force in the model we characterized the dynamics in the conservative case.
In particular, we highlighted the fact that apsidal corotation resonances (ACR) are a powerful tool to understand
the global dynamics of a system.
Then, we showed that the introduction of a dissipative force\footnote{
We consider here tidal effects, but it is clear that other mechanisms could result as well in similar dissipative effects.}
in a resonant system has two main effects.
On the short-term, the system is attracted toward the libration center if it initially relies in its vicinity.
On the long-term, the system tends to follow this ACR solution outside the resonance and both planets
tend to move away from each other.
These two mechanisms are well illustrated and confirmed by the results of two simulations (one for the 2:1 MMR
and the other for the 3:2 MMR)
of the innermost planets of the \object{GJ581} (see section~\ref{sec:simulation}).
Since the ACR solution do not correspond to zero eccentricities even far from the resonance, it is possible to
have resonant angles to oscillate outside of resonances.
However, we showed that in this case the motion is completely characterized by the secular problem and that the fact
that resonant angles appear to librate only means that the secular eigenmodes are (almost) totally damped.
The important fact is that the nature of the motion is the same as when the eigenmodes are not damped
and
\modif{the separatrix of the resonance does not exist anymore in this region of the phase space.}
Thus, it is inappropriate to consider such systems as resonant ones.

\begin{acknowledgements}
We thank Philippe Robutel for helpful discussions.
This work has been supported by PNP-CNRS,
CS of Paris Observatory,
PICS05998 France-Portugal program,
the European Research Council/European Community under the FP7 through a Starting Grant, and
Funda\c{c}\~ao para a Ci\^enca e a Tecnologia, Portugal (grants PTDC/CTE-AST/098528/2008 and PEst-C/CTM/LA0025/2011).
\end{acknowledgements}

\bibliographystyle{aa}
\bibliography{DLCB}

\appendix
\section{Computation of the Hamiltonian coefficients} \label{app:calcoef}
In this section we explain in more details the computations of coefficients $K$, $S$, and $R$ of the
resonant normal form (\refeq{eq:resham}) up to any order.

The Keplerian part is given by:
\begin{equation}
  \mathcal{H}_0 = - \sum_{i=1}^2 \frac{\mu_i^2 \beta_i^3}{2 \Lambda_i^2},
\end{equation}
with:
\begin{eqnarray}
    \Lambda_1 &=& - \frac{p}{q} \left(G + \mathcal{D} - 1 \right),\\
    \Lambda_2 &=& 1 - \left(1+\frac{q}{p}\right) \Lambda_1
      = \left(1+\frac{p}{q}\right) \left(G + \mathcal{D} \right) - \frac{p}{q}.
\end{eqnarray}
$K$ coefficients are thus simply given by the Taylor series of $1/(1+x)^2$ where $x=\mathcal{D}/(G-1)$ for the inner planet
and $x=\mathcal{D}/\left(G-p/(p+q)\right)$ for the outer one.

For the pertubation, the first step is to determine which inequalities $k_1 \lambda_1 + k_2\lambda_2$ have to be computed.
The secular part is given by the inequality $k_1 = k_2 = 0$.
The resonant terms are given by inequalities $k_1 = - k p$, $k_2 = k (p+q)$, with $-d/q\leq k \leq d/q$ where $d$ is the
chosen degree of development.
Each inequality is computed as a power series of $x_1$, $x_2$ by using the algorithm presented in
\citet{laskar_stability_1995}.
For instance, in the case of the 2:1 resonance, the perturbative part of the Hamiltonian reads:
\begin{equation}\label{eq:hamtot}
  \mathcal{H}_1 = -\frac{m_1}{m_0} \Frac{\mu_2^2 \beta_2^3}{\Lambda_2^2} \mathcal{H}_{1,\mathrm{d}}
  + \frac{\mu_1 \beta_1^2}{\Lambda_1} \Frac{\mu_2 \beta_2^2}{\Lambda_2} \mathcal{H}_{1,\mathrm{i}},
\end{equation}
where $\mathcal{H}_{1,\mathrm{d}}$ is the direct part of the development, and
$\mathcal{H}_{1,\mathrm{i}}$ is the indirect part.
The direct part is given by:
\begin{equation}\label{eq:dirH}
  \begin{array}{l c l}
  \mathcal{H}_{1,\mathrm{d}} & = &
      s^{(0)}
      + s_1^{(2)} X_1\overline{X}_1 + s_2^{(2)} X_2\overline{X}_2
      + s_{12}^{(2)} (X_1\overline{X}_2 + X_2\overline{X}_1)\\
      & + & r_1^{(1)} (X_1 + \overline{X}_1) + r_{2,\mathrm{d}}^{(1)} (X_2 + \overline{X}_2)\\
      & + & r_1^{(2)} (X_1^2 + \overline{X}_1^2) + r_2^{(2)} (X_2^2 + \overline{X}_2^2)\\
      & + & r_{12}^{(2)} (X_1 X_2 + \overline{X}_1\overline{X}_2),
  \end{array}
\end{equation}
and the indirect part reads:
\begin{equation}\label{eq:indH}
  \mathcal{H}_{1,\mathrm{i}} = r_{2,\mathrm{i}}^{(1)} (X_2 + \overline{X}_2),
\end{equation}
where we note $X_i = \sqrt{\frac{2}{\scaled{\Lambda}_i}} \scaled{x}_i = \sqrt{\frac{2}{\Lambda_i}} x_i$
\citep[see][]{laskar_stability_1995}.
The coefficients $S$, $R$ of \refeq{eq:resham} correspond to coefficients $s$, $r$
appearing in Eqs.~(\ref{eq:dirH}), (\ref{eq:indH}) but with a renormalization due
to the use of $x_1$, $x_2$ instead of $X_1$, $X_2$ and the factors in front of the direct and indirect parts
in \refeq{eq:hamtot}.
$s$, $r$ coefficients can be expressed as functions of
$\alpha$ and Laplace coefficients $\left(b_s^{(j)}(\alpha)\right)$:
\begin{equation}
\EQM{
s^{(0)} & = & \frac{1}{2} b_{1/2}^{(0)} (\alpha) \crm
s_1^{(2)} & = & \frac{1}{8} \alpha b_{3/2}^{(1)} (\alpha) \crm
s_2^{(2)} & = & \frac{1}{8} \alpha b_{3/2}^{(1)} (\alpha) \crm
s_{12}^{(2)} & = & \frac{3}{8} \alpha b_{3/2}^{(0)} (\alpha)
    - \frac{1}{4} \left(1 + \alpha^{2} \right) b_{3/2}^{(1)} (\alpha)\crm
r_1^{(1)} & = & 
    - \left( \frac{1}{2}
    + \frac{5}{4} \alpha^{2} \right) b_{3/2}^{(0)} (\alpha)\crm
    &&+ \left( \frac{1}{3} \alpha^{-1}
    + \frac{7}{12} \alpha
    + \frac{5}{6} \alpha^{3} \right) b_{3/2}^{(1)} (\alpha)
\crm
r_{2,\mathrm{d}}^{(1)} & = &
    \frac{5}{4} \alpha b_{3/2}^{(0)} (\alpha)
    - \left( \frac{1}{2}
    + \frac{3}{4} \alpha^{2} \right) b_{3/2}^{(1)} (\alpha)\crm
r_{2,\mathrm{i}}^{(1)} & = & \frac{1}{2} \crm
r_1^{(2)} & = & 
    \php\left( \frac{12}{35} \alpha^{-2}
    + \frac{71}{140}
    + \frac{67}{70} \alpha^{2}
    + \frac{138}{35} \alpha^{4} \right) b_{3/2}^{(0)} (\alpha) \crm
    &&- \left( \frac{8}{35} \alpha^{-3}
    + \frac{89}{210} \alpha^{-1}
    + \frac{263}{336} \alpha
    + \frac{341}{210} \alpha^{3}
    + \frac{92}{35} \alpha^{5} \right) b_{3/2}^{(1)} (\alpha)
\crm
r_2^{(2)} & = & 
    \php\left( 1
    + \frac{13}{4} \alpha^{2} \right) b_{3/2}^{(0)} (\alpha) \crm
   && - \left( \frac{2}{3} \alpha^{-1}
    + \frac{65}{48} \alpha
    + \frac{13}{6} \alpha^{3} \right) b_{3/2}^{(1)} (\alpha)
\crm
r_{12}^{(2)} & = & 
    - \left(\frac{6}{5} \alpha^{-1}
    + \frac{69}{40} \alpha
    + \frac{36}{5} \alpha^{3} \right) b_{3/2}^{(0)} (\alpha) \crm
   && + \left( \frac{4}{5} \alpha^{-2}
    + \frac{29}{20}
    + \frac{59}{20} \alpha^{2}
    + \frac{24}{5} \alpha^{4} \right) b_{3/2}^{(1)} (\alpha)
}
\end{equation}

As for the Keplerian part, we substitute $\Lambda_1$ and $\Lambda_2$
by their power series in $\mathcal{D}$ each time they appear in the expression of $\mathcal{H}_1$.
For the Laplace coefficients we first need to develop them in power series around
the nominal resonant semi-major axes ratio which is defined by:
\begin{equation}
  \alpha_r = \left(\frac{\mu_1}{\mu_2}\right)^{1/3} \left(\frac{p}{p+q}\right)^{2/3}.
\end{equation}
Then the deviation $\delta \alpha = \alpha - \alpha_r$ to the nominal value is substituted by its power series in $\mathcal{D}$.
Of course the degrees of development of all these series are adjusted in order
to be consistent with the desired degree in eccentricities ($d$).

\section{Asymptotic evolution of $G$}\label{app:asympt}
In this section we show how to estimate the asymptotic evolution of $G$.
This allows to determine the position of the system in our graphs as a function of time.
We have seen that the evolution of $G$ during the tidal circularization phase
is slower and slower.
Moreover, the escape from the resonance is very quick and the evolution of the system
after this escape is very slow.
Thus, the initial position of the system in the resonance
does not influence much its final position (outside the resonance) after a long time.

The evolution of $G$ during the circularization phase is governed by \refeq{eq:dotG2}.
Since $G$ remains very close to $G_0$, and $\Lambda_i$ stay very close to $\Lambda_{i,r}$,
we can substitute these values in the expression of $\dot{G}$.
However, $e_i$ cannot be considered as constants since the ACR solution tends to zero eccentricities
when $G$ increases.
Thus we have to compute $e_i$ as functions of $\delta G = G - G_0$ with the help of Eqs.~(\ref{eq:acr1}),
(\ref{eq:acr2}).
In the asymptotic evolution, eccentricities are very low and the ACR position can be well approximated with lowest order terms :
\begin{eqnarray}
  0 &\approx& K^{(2)} x_1 + R_1^{(1)},\\
  0 &\approx& K^{(2)} x_2 + R_2^{(1)}.
\end{eqnarray}
Thus the ACR solution is simply given by:
\begin{eqnarray}
  x_1 &=& - \frac{R_1^{(1)}}{ K^{(2)} },\\
  x_2 &=& - \frac{R_2^{(1)}}{ K^{(2)} },
\end{eqnarray}
and $e_i$ are given by:
\begin{equation}
  e_i \approx \sqrt{\Frac{2}{\Lambda_{i,r}}} x_i.
\end{equation}
$K^{(2)}$, $R_1^{(1)}$, and $R_2^{(1)}$ are functions of $G$ but
$R_1^{(1)}$, and $R_2^{(1)}$ are almost constants and can be approximated with their values for $G=G_0$
(see appendix~\ref{app:calcoef}):
\begin{eqnarray}
  R_1^{(1)} &\approx& -\frac{m_1}{m_0} \Frac{\mu_2^2 \beta_2^3}{\Lambda_{2,r}^2}
  \sqrt{\Frac{2}{\Lambda_{1,r}}} r_1^{(1)}(\alpha_r),\\
  R_2^{(1)} &\approx& -\frac{m_1}{m_0} \Frac{\mu_2^2 \beta_2^3}{\Lambda_{2,r}^2}
  \sqrt{\Frac{2}{\Lambda_{2,r}}} r_{2,d}^{(1)}(\alpha_r)\nonumber\\
  &&+ \frac{\mu_1 \beta_1^2}{\Lambda_{1,r}} \Frac{\mu_2 \beta_2^2}{\Lambda_{2,r}}
  \sqrt{\Frac{2}{\Lambda_{2,r}}} r_{2,i}^{(1)}.
\end{eqnarray}
On the contrary, in first approximation $K^{(2)}$ is proportional to $\delta G$:
\begin{equation}
  K^{(2)} \approx -3 \left[ \left(\frac{p}{q}\right)^2 \frac{\mu_1^2 \beta_1^3}{\Lambda_{1,r}^4}
  + \left(1+\frac{p}{q}\right)^2 \frac{\mu_2^2 \beta_2^3}{\Lambda_{2,r}^4} \right] \delta G
\end{equation}
Finally, this means that:
\begin{equation}
  e_i \propto \frac{1}{\delta G},
\end{equation}
and
\begin{equation}
  \delta\dot{G} = \dot{G} \propto \frac{1}{\delta G^2}.
\end{equation}
The asymptotic evolution of $G$ is thus given by:
\begin{equation}
  \delta G \propto t^{1/3}.
\end{equation}
The eccentricities evolve as:
\begin{equation}
  e_i \propto t^{-1/3}.
\end{equation}
The semi-major axes ratio asymptotic evolution is governed by:
\begin{equation}
  \delta \alpha \propto \delta G \propto t^{1/3}.
\end{equation}
We thus find the same asymptotic law ($t^{1/3}$) as \citet{papaloizou_tidal_2011},
and \citet{lithwick_resonant_2012}.

\section{2:1 resonance between proper modes around the ACR solution in $S2$}\label{app:ressec}
\figo
In this section we present a brief analysis of the event happening in $S2$ around $t=1.4\ \mathrm{Myr}$ (Fig.~\ref{Figii}).
We interpret the increase of the amplitude of oscillations in the direction of $e_2$
as a consequence of the crossing of the 2:1 resonance between both proper modes around the ACR solution.
The linear equations of motion around the fixed point (see \refeq{eq:propmod}) correspond to degree two terms
in the diagonalized Hamiltonian:
\begin{equation}
  \mathcal{H}_{\mathrm{diag}, 2} = - g_1 u_1 \overline{u}_1 - g_2 u_2 \overline{u}_2
\end{equation}
The resonance appears in this Hamiltonian at degree three with the term of the form:
\begin{equation}
  \rho \left( u_1 \overline{u}_2^2 + \overline{u}_1 u_2^2 \right),
\end{equation}
where $\rho$ us a constant depending on the parameters and initial conditions.
Let us note $u_i = \sqrt{U_i} \expo{\ImUnit \upsilon_i}$.
We can average over all other angles than the resonant one ($\upsilon_1 - 2\upsilon_2$) with a similar procedure than
the one used in section \ref{sec:conservative}.
The system is then reduced to one degree of freedom and we have a new constant of the (averaged) motion:
\begin{equation}
  U = 2 U_1 + U_2.
\end{equation}
The energy levels curves for the values of the constants $G$ and $U$ taken by the system of simulation $S2$
at the moment of the event (around $t=1.4\ \mathrm{Myr}$) are shown in Fig.~\ref{Figo} superimposed with the trajectory
of the system.
We see that before the event, the proper mode $2$ is almost totally damped compared to
proper mode $1$ (red dots in Fig.~\ref{Figo}),
whereas just after the event, this proper mode gained some amplitude (green dots in Fig.~\ref{Figo}) and the system
follows very well the expected motion.
The resonant angle $\upsilon_1-2\upsilon_2$ does not enter in libration in the simuation $S2$ but this may be due to the
very strong dissipation present in this simulation and which induces an evolution of the constant $G$ quicker than
expected for a real system. Thus the frequencies $g_1$, $g_2$ and the phase space evolve very quickly and the system crosses
the resonance before it has time to enter in the libration area.
The important fact is that due to the crossing of the resonance, the proper mode $2$ increases its amplitude while
the proper mode $1$ amplitude's decreases ($U = 2 U_1 + U_2$ stays constant).
However, since the proper mode $1$ has initially a much higher amplitude, this decrease is imperceptible.
Moreover, since the diagonalizing matrix is dominated by diagonal terms, the evolution of $e_1$ is dominated
by the proper mode $1$ and the evolution of $e_2$ by the proper mode $2$.
Thus, $e_1$ is only weakly affected by the event while $e_2$ is strongly affected.

\end{document}